\documentclass{elsarticle}
\usepackage{amsmath}
\usepackage{multirow}
\usepackage[table]{xcolor}
\usepackage{hyperref}
\usepackage{lmodern}

\usepackage{tabulary}
\usepackage{amssymb}
\usepackage{bm}
\usepackage{graphicx}
\usepackage{wasysym}
\usepackage{url}
\usepackage{algorithm}
\usepackage{algorithmic}
\usepackage{xcolor}
\usepackage{color}
\usepackage{subfig}
\usepackage{enumitem}
\usepackage{mathtools}
\usepackage{comment}
\usepackage{url}

\usepackage{boxedminipage}

\usepackage{ifpdf}
\usepackage{ifthen}

\ifpdf
  \newcommand{\inputfig}[1]{
  }
\else
  \newcommand{\inputfig}[1]{%
\fi

\newcommand{\tsigma}{\mathcal{T}_{\Sigma}}
\newcommand{\tsigmav}{\tsigma(\Var)}
\newcommand{\cryptn}[2]{\{#2\}_{#1}}
\newcommand{\sign}[2]{\mathsf{sign}_{#1}(#2)}
\newcommand{\scrypt}[2]{\{\!| #2 |\!\}_{#1}}
\newcommand{\inv}[1]{\iffont{inv}(#1)}
\newcommand{\invNA}{\iffont{inv}}
\newcommand{\pair}[2]{\langle #1, #2 \rangle}
\newcommand{\fst}[1]{\pi_1(#1)}
\newcommand{\snd}[1]{\pi_2(#1)}

\newcommand{\idfont}[1]{\mathit{#1}}
\newcommand{\HN}{\idfont{HN}}
\newcommand{\NA}{\idfont{NA}}
\newcommand{\NB}{\idfont{NB}}
\newcommand{\na}{\idfont{na}}
\newcommand{\nb}{\idfont{nb}}
\newcommand{\GX}{\idfont{GX}}
\newcommand{\GY}{\idfont{GY}}

\newcommand{\pkNA}{\iffont{pk}}
\newcommand{\skNA}{\iffont{sk}}
\newcommand{\pk}[1]{\pkNA(#1)}
\newcommand{\sk}[1]{\skNA(#1)}
\newcommand{\pkprime}[1]{\iffont{pk'}(#1)}

\newcommand{\senc}{\iffont{enc}}
\newcommand{\sdec}{\iffont{dec}}

\newcommand{\ckNA}{\mathsf{pk}}
\newcommand{\akNA}{\mathsf{sk}}
\newcommand{\ak}[1]{\akNA(#1)}
\newcommand{\ck}[1]{\ckNA(#1)}

\newcommand{\dyS}{\mathcal{DY}}
\newcommand{\dy}[1]{\dyS(#1)}
\newcommand{\dym}[2]{\dyS_{#1}(#2)}
\newcommand{\dymS}[1]{\dyS_{#1}}
\newcommand{\public}{\Sigma_p}

\newcommand{\state}[2]{\stateNA_{#1}(#2)}
\newcommand{\stateNA}{\iffont{state}}
\newcommand{\iknows}[1]{\iknowsNA(#1)}
\newcommand{\iknowsNA}{\iffont{iknows}}
\newcommand{\ifnot}[1]{\iffont{not}(#1)}
\newcommand{\ifdot}{{}_{\;\bullet\;}}

\newcommand{\ifarrow}[1][]{
  \ifthenelse{\equal{#1}{}}{
    \Rightarrow
  }{
    =\!\!\![{#1}]\!\!\!\hspace{1pt}\Rightarrow
  }}

\newcommand{\iffont}[1]{\mathsf{#1}}

\newcommand{\roleA}{\iffont{roleA}} %
\newcommand{\roleB}{\iffont{roleB}} %
\newcommand{\roleR}{\iffont{roleR}} %
\newcommand{\roleC}{\iffont{roleC}} %

\newcommand{\secCh}{\mbox{$\,\bullet\!\!\rightarrow\!\!\bullet\,$}}
\newcommand{\athCh}{\mbox{$\,\bullet\!\!\rightarrow$\,}}
\newcommand{\cnfCh}{\mbox{${\rightarrow\!\!\bullet\,}$}}
\newcommand{\secRCh}{\,\bullet\!\!\!\twoheadrightarrow\!\!\bullet\,}
\newcommand{\athRCh}{\,\bullet\!\!\!\twoheadrightarrow}
\newcommand{\insecCh}{\rightarrow}

\newcommand{\fresh}[1]{\stackrel{\mbox{\scriptsize @}\,}{#1}}
\newcommand{\forward}{\looparrowright}
\newcommand{\fforward}{\fresh{\forward}}
\newcommand{\farrow}{\fresh{\rightarrow}}

\newcommand{\onionCh}{\,\bullet[\rightarrow]\bullet\,}

\newcommand{\idmxCh}{\onionCh}

\newcommand{\dotChdot}{\,\bullet\!\!\leftrightarrow\!\!\bullet\,}
\newcommand{\dotCh}{\,\bullet\!\!\leftrightarrow}
\newcommand{\Chdot}{\leftrightarrow\!\!\bullet\,}

\newcommand{\PsecChP}{\,\circ\!\!\!\rightarrow\!\!\circ\,}
\newcommand{\PsecCh}{\,\circ\!\!\!\rightarrow\!\!\bullet\,}
\newcommand{\secChP}{\,\bullet\!\!\!\rightarrow\!\!\circ\,}
\newcommand{\PathCh}{\,\circ\!\!\!\rightarrow}
\newcommand{\cnfChP}{\rightarrow\!\!\circ\,}

\newcommand{\PChP}{\,\circ\!\!\leftrightarrow\!\!\circ\,}
\newcommand{\dotChP}{\,\bullet\!\!\leftrightarrow\!\!\dot\,}
\newcommand{\PChdot}{\,\circ\!\!\leftrightarrow\!\!\bullet\,}
\newcommand{\PCh}{\,\circ\!\!\leftrightarrow}
\newcommand{\ChP}{\leftrightarrow\!\!\circ\,}

\newcommand{\FathChNA}{\iffont{athCh}}
\newcommand{\FcnfChNA}{\iffont{cnfCh}}
\newcommand{\FsecChNA}{\iffont{secCh}}
\newcommand{\FfathChNA}{\iffont{fathCh}}
\newcommand{\FfsecChNA}{\iffont{fsecCh}}
\newcommand{\FsecCh}[4]{\FsecChNA(#1;#2;#3;#4)}
\newcommand{\FathCh}[3]{\FathChNA(#1;#2;#3)}
\newcommand{\FcnfCh}[2]{\FcnfChNA(#1;#2)}
\newcommand{\FfathCh}[4]{\FfathChNA^{#1}(#2;#3;#4)}
\newcommand{\FfsecCh}[5]{\FfsecChNA^{#1}(#2;#3;#4;#5)}

\newcommand{\athIssue}[1]{\iffont{athIssue(#1)}}
\newcommand{\cnfIssue}[1]{\iffont{cnfIssue(#1)}}
\newcommand{\secIssue}[1]{\iffont{secIssue(#1)}}
\newcommand{\athRely}[1]{\iffont{athRely(#1)}}
\newcommand{\cnfRely}[1]{\iffont{cnfRely(#1)}}
\newcommand{\secRely}[1]{\iffont{secRely(#1)}}

\newcommand{\pkEnc}[1]{\mathit{pkEnc}(#1)}
\newcommand{\pkSig}[1]{\mathit{pkSig}(#1)}

\newcommand{\DownGrade}{\mathrm{DownGrade}}
\newcommand{\Combine}{\mathrm{Combine}}
\newcommand{\SymOne}{\mathrm{Sym_1}}
\newcommand{\SymTwo}{\mathrm{Sym_2}}
\newcommand{\SymTre}{\mathrm{Sym_3}}
\newcommand{\CreatePseudo}{\mathrm{CreatePseudo}}
\newcommand{\UsePseudo}{\mathrm{UsePseudo}}
\newcommand{\UseRealName}{\mathrm{UseRealName}}
\newcommand{\PseudoDownGrade}{\mathrm{PseudoDownGrade}}
\newcommand{\AuthPseudo}{\mathrm{AuthPseudo}}
\newcommand{\AthTTP}{\mathrm{AthTTP}}
\newcommand{\CnfTTP}{\mathrm{CnfTTP}}
\newcommand{\SecTTP}{\mathrm{SecTTP}}

\newcommand{\honest}[1]{\mathit{honest}(#1)}
\newcommand{\dishonest}[1]{\iffont{dishonest}(#1)}
\newcommand{\vddash}{\vdash\!\!\!\vdash}
\newcommand{\mmodels}{\models\hspace{-0.3cm}\models}

\newcommand{\mode}[3]{({#1} \,|\, {#2} \,|\, {#3})}

\newcommand{\IK}{\mathit{IK}}
\newcommand{\agent}[1]{\iffont{agent}(#1)}

\newcommand{\Fresh}{\mathit{Fresh}}
\newcommand{\Payload}{\mathit{Payload}}
\newcommand{\Public}{\mathit{Public}}
\newcommand{\Tag}{\mathit{Tag}}
\newcommand{\lift}[1]{\lceil #1\rceil}

\newcommand{\pos}[1]{\mathit{pos}(#1)}
\newcommand{\Var}{\mathcal{V}}

\newcommand{\tagfont}[1]{\mathsf{#1}}
\newcommand{\tSone}{\tagfont{S_1}}
\newcommand{\tStwo}{\tagfont{S_2}}

\newcommand{\optfix}[2]{}
\newcommand{\dyM}[2]{\mathcal{DY}_{#1}(#2)}
\newcommand{\nf}[1]{#1_{\downarrow C/F}}
\newcommand{\pattern}[2]{{\lceil\,#1\,\rceil}_{#2}}
\newcommand{\decryptPat}[2]{{\lceil\!\!\lceil\,#1\,\rceil\!\!\rceil}_{#2}}

\newcommand{\pubNA}{\mathit{pub}}
\newcommand{\responseNA}{\mathit{response}}
\newcommand{\checkRNA}{\mathit{check}}

\newcommand{\pub}[1]{\pubNA(#1)}
\newcommand{\response}[1]{\responseNA(#1)}
\newcommand{\checkR}[1]{\checkRNA(#1)}

\newcommand{\Keyword}[1]{\mathsf{#1}}

\newcommand{\Protocol}[5]{
  \Keyword{Protocol:}~\mathit{#1}\\
  \Keyword{Types:}\\
  #2
  \Keyword{Knowledge:}\\
  #3
  \Keyword{Actions:}\\
  \begin{array}{cllllcl}
  #4
  \end{array}\\
  \Keyword{Goals:}\\
  \begin{array}{ccccl}
  #5
  \end{array}
}
\newcommand{\MidProtocol}[5]{
  \begin{array}{ccccl}
  #4
  \end{array}\\
}

\newcommand{\MSC}[5]{
  \Keyword{Protocol:}~\mathit{#1}\\
  \begin{array}{ccccl}
    #4
  \end{array}
}

\newcommand{\ShortProtocol}[5]{
  \Keyword{Protocol:}~\mathit{#1}\\
  \begin{array}{ccccl}
  #4
\end{array}\\
}
\newcommand{\CompactProtocol}[5]{
  \begin{array}{ccccl}
  #4
  \hline
  #5
  \end{array}\\
}
\newcommand{\Type}[2]{\quad #1\;\mathit{#2};\\}
\newcommand{\Agent}{\Keyword{Agent}}
\newcommand{\Number}{\Keyword{Number}}
\newcommand{\Function}{\Keyword{Function}}
\newcommand{\TFunction}{\Keyword{Function}}
\newcommand{\Knowledge}[2]{\quad\mathit{#1}:~\mathit{#2};\\}
\newcommand{\Create}[2]{
  \multicolumn{5}{l}{\quad\#\mathit{#1}~\text{creates}~\mathit{#2}}\\}
\newcommand{\Let}[2]{
  \multicolumn{5}{l}{\quad\#\mathit{#1}~:=~\mathit{#2}}\\}
\newcommand{\Action}[5]{\quad\mathit{#1}#2\mathit{#3},\mathit{#4}:&\mathit{#5}\\}
\newcommand{\Repeat}[5]{\multicolumn{5}{l}{
    \quad#1=\mathit{#2}#3\mathit{#4}:\mathit{#5}}\\}
\newcommand{\NGoal }[4]{\quad\mathit{#1}&#2&\mathit{#3}&:&\mathit{#4}\\}
\newcommand{\AuthGoal}[3]
{\multicolumn{5}{l}{
    \quad\mathit{#1}~\Keyword{authenticates}~\mathit{#2}~%
    \Keyword{on}~\mathit{#3}}\\}
\newcommand{\SecGoal}[2]
{\multicolumn{5}{l}{\quad\mathit{#1}~\Keyword{secret~between}~\mathit{#2}}\\}
\newcommand{\REML}[1]{\multicolumn{5}{l}{\quad\#\text{ #1}}\\}

\newcommand{\ifrule}[3]{#1\\ \ifarrow[#2]\\ #3\\[2ex]}

\newcommand{\subterm}{\sqsubseteq}
\newcommand{\propersubterm}{\sqsubset}
\newcommand{\supterm}{\sqsupseteq}
\newcommand{\propersupterm}{\sqsupset}

\newcommand{\decryptions}[2]{\mathit{decryptions}_{#1}(#2)}
\newcommand{\patternR}[3]{\pattern{#2}{#3}^{#1}}

\newcommand{\instanceof}{\succeq}

\newcommand{\secret}[2]{\iffont{secret}(#1,#2)}

\newcommand{\idemix}{\textsf{Identity Mixer}}

\newcommand{\xor}{\oplus}
\newcommand{\algo}[1]{\ensuremath{\mathsf{#1}}}
\newcommand{\const}[1]{\algo{#1}}
\newcommand{\vari}[1]{\ensuremath{\mathit{#1}}}

\newcommand{\ana}[2]{\mathit{ana}_{#1}(#2)}
\newcommand{\derivations}[3]{\mathit{derivations}_{#1}(#2,#3)}
\newcommand{\dereq}[3]{\mathit{dereq}_{#1}(#2,#3)}
\newcommand{\checks}[3]{\mathit{checks}_{#1}(#2,#3)}
\newcommand{\given}[1]{\textsl{given}(#1)}
\newcommand{\ungive}[1]{{#1}_*}

\definecolor{grey}{rgb}{0.5,0.5,0.5}\newcommand{\grey}{\color{grey}}
\definecolor{red}{rgb}{1,0,0}\newcommand{\red}{\color{red}}
\definecolor{green}{rgb}{0,0.45,0}\newcommand{\green}{\color{green}}
\definecolor{blue}{rgb}{0,0,1}\newcommand{\blue}{\color{blue}}

\newcommand{\rpif}{\mathbb{P}}
\newcommand{\rif}{\mathbb{R}}
\newcommand{\rf}{\mathbb{F}}
\newcommand{\re}{\mathbb{E}}
\newcommand{\md}{\mathbb{M}}
\newcommand{\traces}{\mathbb{T}}
\newcommand{\sigS}{\iffont{pk}(S)}
\newcommand{\send}[2]{\iffont{snd}_{#1}(#2)}
\newcommand{\recv}[2]{\iffont{rcv}_{#1}(#2)}
\newcommand{\trace}{t}
\newcommand{\evs}{\mathit{evs}}
\newcommand{\evt}{\mathit{evt}}
\newcommand{\mkset}[1]{[#1]} %
\newcommand{\player}[1]{\mathit{player}(#1)}
\newcommand{\used}[1]{{\mathit{ used\;#1}}}

\newcommand{\sem}[1]{[\![ #1 ]\!]}

\newcommand{\mX}{\mathcal{X}}

\newcommand{\PVar}{\mathcal{P}}
\newcommand{\SigmaP}{\Sigma_\PVar}
\newcommand{\tsigmap}{\mathcal{T}_{\SigmaP}}
\newcommand{\arity}[1]{\mathit{arity}(#1)}

\newcommand{\Nat}{\mathbb{N}}
\newcommand{\dcrypt}[2]{\cryptn{#1}{#2}^{-1}}
\newcommand{\dscrypt}[2]{\scrypt{#1}{#2}^{-1}}
\newcommand{\ccs}[1]{\mathit{ccs}(#1)}

\newcommand{\interpretation}{\mathcal{I}}
\newcommand{\intruder}{\mathsf{i}}

\newcommand{\know}[1]{\mathit{know}(#1)}
\newcommand{\verify}[1]{\mathit{verify}(#1)}
\newcommand{\true}{\mathit{true}}

\newcommand{\anl}[1]{&&&&\%\mathit{#1}\\}
\newcommand{\REM}[1]{\;\;\Keyword{\#}\;\;\text{#1}}

\newcommand{\StoreOA}[2]{\Keyword{Store}_{#1}(\,#2\,)}
\newcommand{\CheckStoreOA}[2]{\Keyword{CheckStore}_{#1}(\,#2\,)}
\newcommand{\LoadOA}[2]{#2\leftarrow\Keyword{Load}_{#1}}

\newcommand{\Store}[2]{\multicolumn{5}{l}{
    \quad\StoreOA{#1}{#2}}\\}
\newcommand{\Load}[2]{\multicolumn{5}{l}{
    \quad\LoadOA{#1}{#2}}\\}
\newcommand{\CheckStore}[2]{\multicolumn{5}{l}{
    \quad\CheckStoreOA{#1}{#2}}\\}

\newcommand{\FormNym}{\mathsf{FormNym}}
\newcommand{\GrantCred}{\mathsf{GrantCred}}
\newcommand{\VerifyCred}{\mathsf{VerifyCred}}
\newcommand{\VerifyCredOnNym}{\mathsf{VerifyCredOnNym}}
\newcommand{\masecNA}{\iffont{masec}}
\newcommand{\masec}[1]{\masecNA(#1)}
\newcommand{\ptagNA}{\iffont{p}}
\newcommand{\ptag}[1]{\ptagNA(#1)}

\newcommand{\commitNA}{\iffont{commit}}
\newcommand{\commitIINA}{\iffont{commit_2}}
\newcommand{\commitIIINA}{\iffont{commit_3}}

\newcommand{\commit}[1]{\commitNA(#1)}
\newcommand{\commitII}[1]{\commitIINA(#1)}
\newcommand{\commitIII}[1]{\commitIIINA(#1)}

\newcommand{\zkpNA}{\iffont{zkp}}
\newcommand{\user}[1]{\iffont{user}(#1)}
\newcommand{\owner}[1]{\iffont{owner}(#1)}

\newcommand{\Always}{\iffont{Always}}
\newcommand{\Previously}{\iffont{Previously}}

\newcommand{\pending}[1]{\iffont{pending}(#1)}

\newcommand{\pack}[1]{\mathit{pack}(#1)}
\newcommand{\hide}[1]{\mathit{hide}(#1)}

\newcommand{\ToTrusted}[1]{\Action{#1}{\idmxCh}{T}}
\newcommand{\TrustedTo}[1]{\Action{T}{\idmxCh}{#1}}

\newcommand{\registered}[1]{\iffont{registered}(#1)}
\newcommand{\granted}[1]{\iffont{granted}(#1)}

\newcommand{\verified}[1]{\iffont{verified}(#1)}

\newcommand{\dhh}{\mathit{dhh}}
\newcommand{\dhf}{\mathit{dhf}}
\newcommand{\DHtag}[1]{\mathit{tag}(#1)}
\newcommand{\hk}{\mathit{hk}}
\newcommand{\fk}{\mathit{fk}}
\newcommand{\exps}{\mathit{exps}}

\newcommand{\from}[2]{\fromnoarg(#1;#2)}
\newcommand{\fromnoarg}{\mathit{from}}

\renewcommand{\vector}[1]{\overline{#1}}

\newcommand{\vars}[1]{\mathit{vars}(#1)}
\newcommand{\dom}[1]{\mathit{dom}(#1)}

\newcommand{\drightarrow}{\rightarrow\!\!\!\!\!\rightarrow}
\newcommand{\bigsem}[1]{\left[\!\!\left[#1\right]\!\!\right]}

\newcommand{\hornarrow}{\rightarrow}
\newcommand{\iknowsa}[1]{\mathsf{iknows}(#1}
\newcommand{\signa}[1]{\mathsf{sign}_{#1}}
\newcommand{\todo}[1]{\fbox{\textbf{TODO: }#1}}
\newcommand{\valid}{\mathit{valid}}
\newcommand{\revoked}{\textit{revoked}}
\newcommand{\db}{\mathit{db}}
\newcommand{\ring}{\mathit{ring}}
\newcommand{\PK}{\mathit{PK}}
\newcommand{\NPK}{\mathit{NPK}}
\newcommand{\occurs}{\mathit{occurs}}
\newcommand{\timplies}{\mathit{implies}}
\newcommand{\LFP}{\mathit{LFP}}

\newcommand{\Habs}{\hat{\mathfrak{A}}}
\newcommand{\Abs}{\mathfrak{A}}
\newcommand{\Vara}{\Var_{\mathfrak{A}}}
\newcommand{\tabs}{\mathcal{T}_\mathfrak{A}}
\newcommand{\thabs}{\hat{\mathcal{T}_\mathfrak{A}}}
\newcommand{\ITIF}{{\it IF}}
\newcommand{\CryptoIF}{{\it CCM}}
\newcommand{\ChanIF}{{\it ICM}}

\newcommand{\ann}{\textit{ann}}
\newcommand{\atag}{\mathsf{atag}}
\newcommand{\fatag}{\mathsf{fatag}}
\newcommand{\ctag}{\mathsf{ctag}}
\newcommand{\stag}{\mathsf{stag}}
\newcommand{\fstag}{\mathsf{fstag}}
\newcommand{\btag}{\mathsf{plain}}
\newcommand{\ftag}{\mathsf{fresh}}
\newcommand{\fotag}{\mathsf{forw}}
\newcommand{\blind}{\mathsf{blind}}
\newcommand{\verif}[1]{#1}
\newcommand{\seenNA}{\mathsf{seen}}
\newcommand{\seen}[1]{\seenNA(#1)}
\newcommand{\notseen}[1]{\mathsf{fresh}(#1)}

\newcommand{\ttbracket}[1]{\texttt{[}#1\texttt{]}}
\newcommand{\anb}{\textit{AnB}}
\newcommand{\anbx}{\textit{AnBx}}
\newcommand{\IM}{\texttt{IM}}
\newcommand{\CM}{\texttt{CM}}
\newcommand{\trans}[3]{\llbracket#1\rrbracket^{#2}_{#3}}
\newcommand{\enc}[1]{\llbracket #1 \rrbracket}
\newcommand{\itrans}[3]{\llparenthesis #1 \rrparenthesis^{#2}_{#3}}
\newcommand{\after}[2]{#1 \texttt{ after } #2}
\newcommand{\arr}[1]{=\!\![#1]\!\!\Rightarrow}
\newcommand{\x}{{\cal X}}
\newcommand{\y}{{\cal Y}}
\newcommand{\X}{{\cal X}}
\newcommand{\T}{{\cal T}}
\newcommand{\V}{{\cal V}}
\newcommand{\chan}{\textsf{chan}}

\newcommand{\cchannel}{\mathsf{Chan_{CCM}}}
\newcommand{\ichannel}{\mathsf{Chan_{ICM}}}
\newcommand{\mgu}{\textit{mgu}}
\newcommand{\bl}{\mathsf{b}}

\renewcommand{\exp}{\iffont{exp}}

\renewcommand{\dscrypt}[2]{\scrypt{#1}{#2}}
\renewcommand{\dcrypt}[2]{\cryptn{#1}{#2}}

\newcommand{\signed}[2]{\mathsf{S}_{#1}(#2)}
\newcommand{\nonce}[1]{n_{#1}}
\newcommand{\newkey}[1]{\mathsf{new}\enskip{#1}}
\newcommand{\hmac}[2]{\mathsf{hmac}_{#1}(#2)}
\newcommand{\hashsymbol}{\mathcal{H}}
\newcommand{\hash}[1]{\mathsf{hash}(#1)}
\newcommand{\verdigest}[2]{[\textit{#1:#2}]}
\newcommand{\digest}[1]{[\textit{#1}]}

\newcommand{\contains}[2]{\iffont{contains}(#1,#2)}
\newcommand{\error}{\iffont{error}}
\newcommand{\wrequest}[1]{\iffont{wrequest}(#1)}
\newcommand{\request}[1]{\iffont{request}(#1)}
\newcommand{\witness}[1]{\iffont{witness}(#1)}

\newcommand{\maxd}[1]{\mathit{maxd}(#1)}
\newcommand{\sccs}[1]{\mathit{sccs}(#1)}
\newcommand{\nts}[1]{\mathit{#1}}

\newcommand{\channel}[2]{\mathsf{channel}_{#1}(#2)}
\newcommand{\GYS}{\mathit{GYS}}

\newcommand{\ccsF}[1]{\mathit{ccs}_F(#1)}

\newcommand{\olds}[1]{\oldstylenums{#1}}
\newcommand{\oldsb}[1]{{\bfseries\olds{#1}}}
\newcommand{\mnote}[1]{\stepcounter{ncomm}%
\vbox to0pt{\vss\llap{\tiny\oldsb{\arabic{ncomm}}}\vskip6pt}%
\marginpar{\tiny\bf\raggedright%
{\oldsb{\arabic{ncomm}}}.\hskip0.5em#1}}\newcounter{ncomm}

\newcommand{\miknote}[1]{{\red \mnote{{\red m: #1}}}}

\usepackage{versions}
\excludeversion{abstractionslearningextra}
\usepackage{tikz}
\usetikzlibrary{shapes,arrows,matrix,backgrounds,shapes,positioning,petri,topaths}
\usetikzlibrary{chains,calc}
\usetikzlibrary{decorations.pathreplacing}
\setcounter{secnumdepth}{3}
\usepackage{pgfplots}
\pgfplotsset{compat=1.9}
\usepackage{booktabs}
\usepackage{hyperref}
\usepackage{doi}
\usepackage{pgf-umlsd}
\usepackage{smartdiagram}
\usepackage{rotating}
\usepackage{longtable}
\usepackage{nomencl}

\makenomenclature
\renewcommand{\nomname}{}
\setlength{\nomitemsep}{0.25mm}

\includeversion{SHORT}
\excludeversion{LONG}

\renewcommand{\ttdefault}{pcr}

\colorlet{punct}{red!60!black}
\definecolor{background}{HTML}{EEEEEE}
\definecolor{delim}{RGB}{20,105,176}
\definecolor{darkgreen}{RGB}{1,50,32}
\colorlet{numb}{magenta!60!black}

\definecolor{pgreen}{rgb}{0,0.5,0}

\usepackage{listings}
\lstdefinelanguage{AnBx}{
	keywords = {Protocol:, Types:, Agent, Number, Function, Certified, SymmetricKey, PublicKey, Untyped, UUID, JWT, IntentNo, AccountNumber, Definitions:, Knowledge:, where, agree, share, Actions:, Goals:, weakly, authenticates, on, secret, between, empty},
	keywordstyle=\color{blue},
	keywordstyle=[2]{\color{red}},
	alsoletter=:,
	breaklines=true,
	basicstyle=\footnotesize\ttfamily,%
	keywordstyle=\bfseries,
	commentstyle=\color{pgreen}\itshape,
	morecomment=[l]{\#}
}

\lstdefinelanguage{OFMC}{
    breaklines=true,
    basicstyle=\ttfamily,
    commentstyle=\color{darkgreen},
    morecomment=[l]{\#}
}

\lstdefinelanguage{json}{
    basicstyle=\tiny\ttfamily,
    stepnumber=1,
    numbersep=8pt,
    showstringspaces=false,
    breaklines=true,
    frame=lines,
    literate=
     *{0}{{{\color{numb}0}}}{1}
      {1}{{{\color{numb}1}}}{1}
      {2}{{{\color{numb}2}}}{1}
      {3}{{{\color{numb}3}}}{1}
      {4}{{{\color{numb}4}}}{1}
      {5}{{{\color{numb}5}}}{1}
      {6}{{{\color{numb}6}}}{1}
      {7}{{{\color{numb}7}}}{1}
      {8}{{{\color{numb}8}}}{1}
      {9}{{{\color{numb}9}}}{1}
      {:}{{{\color{punct}{:}}}}{1}
      {,}{{{\color{punct}{,}}}}{1}
      {\{}{{{\color{delim}{\{}}}}{1}
      {\}}{{{\color{delim}{\}}}}}{1}
      {[}{{{\color{delim}{[}}}}{1}
      {]}{{{\color{delim}{]}}}}{1},
}

\newcommand{\parahead}[1]{\bigskip\noindent\textbf{#1}}

\newcommand{\smallurl}[1]{\footnotesize\url{#1}\normalsize}

\newcommand{\mycomment}[3]{[\textcolor{red}{\textit{#1 $\rightarrow$ #2}}~\textcolor{blue}{:~#3}]}
\newcommand{\lfcomment}[1]{\mycomment{LF}{PM}{#1}}

\setlength{\belowcaptionskip}{-10pt}

\newif\ifshowfixes \showfixestrue
\newcommand{\fix}[2]{\ifshowfixes {\bf FIX}\footnote{{\bf FIX #1}: #2} \else \relax \fi}
\newcommand{\rulesep}{\unskip\ \vrule\ }

\makeatletter
\def\ps@pprintTitle{%
  \let\@oddhead\@empty
  \let\@evenhead\@empty  \let\@oddfoot\@empty
  \let\@evenfoot\@oddfoot
}
\makeatother
\begin{document}
\sloppy
\title{Security Analysis of the Open Banking\\Account and Transaction API Protocol}

\author[1]{Paolo Modesti}
\ead{p.modesti@tees.ac.uk}

\author[2]{Leo Freitas}
\ead{leo.freitas@ncl.ac.uk}

\author[1]{Qudus Shotomiwa}
\ead{b1636952@tees.ac.uk}

\author[3]{Abdulaziz Almehrej}
\ead{a.almehrej@outlook.com}

\address[1]{Departement of Computing and Games, Teesside University, Middlesbrough, UK}
\address[2]{School of Computing, Newcastle University, UK}
\address[3]{Independent researcher; former School of Computing, Newcastle University, UK}

\begin{abstract}

The Second Payment Services Directive (PSD2) of the European Union aims to create a consumer-friendly financial market by mandating secure and standardised data sharing between banking operators and third parties. Consequently, EU countries and the United Kingdom have adopted Open Banking, a standardised data-sharing API. This paper presents a formal modelling and security analysis of the UK Open Banking Standard's APIs, with a specific focus on the Account and Transaction API protocol. Our methodology employs the extended Alice and Bob notation (AnBx) to create a formal model of the protocol, which is then  verified using the OFMC symbolic model checker and the Proverif cryptographic protocol verifier. We extend previous work by enabling verification for unlimited sessions with a strongly typed model. Additionally, we integrate our formal analysis with practical security testing of some necessary conditions to demonstrate verified security-goals in the NatWest Open Banking sandbox, evaluating mechanisms such as authorisation and authentication procedures.

\end{abstract}
\maketitle

\section{Introduction}
\label{sec:introducation}

Limited competition in the financial services industry was a key factor that led the European Union to introduce the second version of the Payment Services Directive (PSD2)~\cite{article:PSD2_directive}. This legislation aims to enhance competition by facilitating and encouraging bank account holders to share their account data in a controlled and secure manner. Alongside economic opportunities \cite{stefanelli2023digital}, this approach entails clear and significant privacy and security implications that must be thoughtfully considered when constructing systems that allow data sharing on such a scale~\cite{Frei2023}.

To establish a standard API for the sharing of customer data across different banks, the UK, akin to other European countries, introduced the Open Banking Standard~\cite{DePascalis2024}. The regulation encompasses several API specifications suitable for various \emph{Third Party Providers} (TPPs) who aim to serve consumers consenting to share their data. The adoption of a standardised interface promotes interoperability and simplifies the implementation of systems for sharing data between banks and TPPs.

While allowing TPPs to drive changes in the financial services industry can stimulate innovative solutions \cite{Ziegler2021}, it is imperative to consider that software failures and vulnerabilities may incur significant costs for organisations. In fact, interoperability between financial institutions can be a lucrative target for cyber attacks~\cite{Achim2024}. For instance, in 2017, US\$$6$ million was stolen from a Russian bank via the SWIFT system~\cite{website:Reuters}. Non-compliance with regulations can also have substantial consequences. In 2016, Tesco Bank was fined \pounds $16.4$ million in the UK for neglecting its responsibility to protect customers from a cyber attack that resulted in the loss of \pounds $2.26$ million~\cite{website:FCA_Tesco_attack}.

The adoption of the EU General Data Protection Regulation (GDPR) 2016/679 \cite{article:GDPR_directive} in 2018 has promoted stronger data protection measures in EU/EEA member states, requiring organisations to adopt appropriate technical and organisational measures to implement the principles enshrined in such legislation. Notably, for severe violations listed in Art. 83(5) GDPR, fines can be up to \euro{}20 million or up to 4\% of the organisation's total global turnover, whichever is higher.

In this context, formally validating complex software systems, such as the one discussed in this work, is crucial to ensure that the system functions correctly according to its specifications and meets a range of desired properties, including explicitly stated security goals. Gounari et al. \cite{gounari2024harmonizing} have identified the lack of clarity in the technical security specifications necessary for the effective implementation of Open Banking as a major challenge. Additionally, Kassab and Laplante \cite{Kassab2022} highlight how trust is essential for consumer acceptance of Open Banking services.

\paragraph{Contribution}

This paper provides a formal modelling and security analysis of the Open Banking Standard APIs, with a specific focus on automatically verifying the Account and Transaction API protocol used by UK banks. Our contributions include:

\begin{itemize}[noitemsep]
	\item \textit{Formal Modelling and Verification}: We use the extended Alice and Bob notation (\textit{AnBx})~\cite{DBLP:journals/istr/BugliesiCMM16} to create a formal model of the protocol, as detailed in Section \ref{sec:implementation}. We formalise various security goals derived from the requirements and verify them against the protocol formal model (Section~\ref{sec:security}). The \textit{AnBx} model is translated into \textit{AnB} notation~\cite{anb} using the \textit{AnBx} compiler~\cite{anbx2015}, and verified with the OFMC symbolic model checker~\cite{basin2005ofmc} for single-session scenarios. Additionally, the \textit{AnBx} compiler produces a typed applied-pi model for verification with ProVerif~\cite{Blanchet2022}, which confirms that security properties hold for an unbounded number of sessions.
	This approach addresses the limitations of previous work~\cite{abz2020}, where verification was restricted to a single session. This is problematic because relevant attacks may involve multiple parallel sessions. Moreover, in the previous attempt, some functions that return random values were modelled as deterministic functions, leading to an over-approximation that resulted in spurious attack traces. %

 \item \textit{Security Testing on a Sandbox Implementation}: In addition to formal analysis, we perform security testing in the NatWest Open Banking sandbox. This test assesses key API security mechanisms, including authorisation and authentication procedures (Section~\ref{sec:sandbox-testing}).

\item \textit{Literature review:~} we thoroughly research available literature on the Open Banking protocol, the related technologies,  and discuss the current state-of the art (Section~\ref{sec:relatedwork}).
\end{itemize}

\section{Background}
\label{sec:background}
The Open Banking APIs allow third-party developers to build applications and services around financial institutions. These APIs allow secure access to financial data and provide opportunities for creating innovative financial solutions, improving customer experiences, and increasing competition within the financial services industry \cite{DePascalis2024,Arner2022,Borgogno2021}.

\subsection{The Open Banking Standard Overview}

The Payment Services Directive (PSD) is the European legislation aimed at enhancing payment services across the EU in terms of safety and innovation~\cite{website:PSD_FAQ}. The initial version of PSD was adopted in 2007~\cite{article:PSD1_directive}. However, as the digitisation of the economy progressed and new payment services emerged, it became outdated and insufficient to ensure consumer protection and adequate competition. Consequently, a revised version was adopted in November 2015 with the goals of: 1) Ensuring that all payment service providers have equal operating conditions; 2) Expanding the market for new means of payment; 3) Ensuring the security of consumers using payment services~\cite{article:PSD2_directive}.

To implement the aims of the directive in the UK, the Open Banking Implementation Entity (OBIE) was established in 2016 by the UK Competition and Markets Authority (CMA). The Open Banking Working Group (OBWG) published a report on the Open Banking Standard~\cite{report:OB_standard}, outlining two key outcomes:

\begin{itemize}[noitemsep]
	\item An open API for sharing data regarding the services offered by \emph{Account Servicing Payment Service Providers} (ASPSPs), such as banks;
	\item An open API for sharing the account data of \emph{Payment Service Users} (PSUs) provided by ASPSPs. These are the end-users consuming the PSD2 services via TPPs.
\end{itemize}

Open Banking identifies not only the API endpoints (such as the location of resources accessible by third parties, including developers, to build banking and financial applications), but also encompasses data and security standards. The \emph{data standard} defines data models for the API data format, the \emph{API standard} outlines operational requirements for the API, and the \emph{security standard} establishes API security requirements \cite{Ziegler2021}.

By June 2023, the standard was adopted by 151 regulated ASPSPs with 208 TPPs providing services~\cite{website:OB_stats}.

This illustrates the significance of Open Banking for the financial services industry~\cite{Borgogno2021} and, consequently, the importance of verifying the standard's correctness. Additionally, the relevance of the UK version of the standard is due to the fact that it represents the most developed Open Banking market. In fact, in 2023, the total number of API calls in the UK was 14 billion, compared to a combined 6.4 billion in Germany, Italy, France and Spain, according to PwC estimates
\cite{website:OB_stats-2024}. %

\subsection{Account and Transaction API Protocol}
\label{sub:atp}

An \emph{Account Information Service Provider} (AISP) is a regulated entity allowed by ASPSPs to access a PSU's account data if the PSU provides their consent. This type of access is read-only as AISPs are not expected to directly affect the payment accounts they are allowed access to. An AISP can then provide different services with the PSU's account and transaction data, including applications that offer a user-friendly view of the status of the various payment accounts held by the PSU, budgeting advice, price comparisons, and product recommendations.

The Account and Transaction API protocol (ATP) flow~\cite{website:OB_AIS}, which allows AISPs access to the PSU's account data, is shown in Figure \ref{fig:AIS_flow}.

\tikzstyle{startstop} = [rectangle, rounded corners, minimum width=3cm, minimum height=1cm, text centered, draw=black, fill=green!20, text width=3.5cm, align=center]
\tikzstyle{process} = [rectangle, minimum width=3cm, minimum height=1cm, text centered, draw=black, fill=green!20, text width=3cm, align=center]
\tikzstyle{arrow} = [->,>=stealth]
\tikzstyle{block} = [rectangle, draw, fill=blue!20, text width=5em, text centered, rounded corners, minimum height=4em]
\tikzstyle{innerblock} = [rectangle, draw, fill=blue!10, text width=6.5em, text centered, rounded corners, minimum height=2em]
\begin{figure}[ht]
	\centering
\begin{tikzpicture}[node distance=2.0cm]
	\scriptsize
	\node (psu) [block] {\textbf{PSU} \\ (Customer)};
	\node (aisp) [block, right of=psu, xshift=4cm] {\textbf{AISP}};
	
	\matrix (aspsp) [draw, fill=blue!20, text centered, below of=psu, yshift=-1cm, xshift=3cm, rounded corners] {
		\node (aspsp-text) {\textbf{ASPSP}}; \\
		\node (res) [innerblock] {Resource\\ Server}; &
		\node (auth) [innerblock] {Authorisation\\ Server}; \\
	};
	
	\node (step1) [startstop, above of=psu, yshift=-1.3cm, xshift=3cm] {\textbf{Step 1}: PSU requests payment account information from AISP};
	\node (step2) [startstop, below of=aisp, yshift=0.5cm, xshift=1.1cm] {\textbf{Step 2}: AISP requests creation of an unauthorised account access consent from PSU's ASPSP};
	\node (step3) [startstop, below of=psu, yshift=0.3cm, xshift=-0.8cm] {\textbf{Step 3}: PSU authorises account access consent};
	\node (step4) [startstop, below of=aisp, yshift=-1cm, xshift= 1.1cm] {\textbf{Step 4}: AISP requests account information from ASPSP};
	
	\draw [arrow] (psu) -- (aisp);		%
	\draw [arrow] (aisp) -- (aspsp);	%
	\draw [arrow] (psu) -- (aspsp);		%

\end{tikzpicture}
\caption{The Account and Transaction API protocol high-level flow}
\label{fig:AIS_flow}
\end{figure}

\begin{description}
	\item[Step 1] The API is initiated with the PSU asking for information regarding their payment account(s) from an AISP.
	\item[Step 2] The AISP then attempts to create an \emph{account access consent} with the corresponding ASPSP, based on the access permissions agreed with the PSU. To do this, the AISP first authenticates itself with the ASPSP through a \emph{client credential grant}, which is an approach to machine-to-machine authentication. The ASPSP then provides the AISP with an \emph{access token}, allowing them to create the consent by subsequently requesting a POST request to the \verb|account-access-consents| endpoint on the ASPSP's resource server. Upon receiving the request, the server creates an \emph{authorised account access consent} with a unique id (\emph{ConsentId}) that is sent back to the AISP. The account access consent data model is as follows:
	\begin{itemize}
		\item \emph{Permissions} -- an array containing the type of data the PSU has granted the AISP access to. This has to include at least the permission to access the basic account details.
		\item \emph{Expiration Date} -- a future date upon which the AISP will no longer have access to the data (optional parameter).
		\item \emph{Transaction Validity Period} -- two (optional) fields indicating the period within which any transactions that have occurred will be accessible by the AISP. The start and end of a statement history will have to be within the specified period to be accessible.
	\end{itemize}
	\item[Step 3] At this point, the created account access consent has to be authorised to be used by the AISP to access the PSU's account data. This requires the PSUs to authenticate themselves to the ASPSP, which can be done in one of two ways, and then authorise the account access consent. The two authentication approaches or grant types are briefly explained next:
	\begin{itemize}\label{aisp_grant-types}
		\item \emph{Authorisation Code Grant} -- the PSU is redirected by the AISP to the ASPSP to authenticate themselves and authorise the account access consent. Subsequently, the PSU is redirected back to the AISP by the ASPSP with an authorisation code. The AISP then contacts the ASPSP's authorisation server to exchange the authorisation code for the account data access token.
		\item \emph{Client Initiated Backchannel Authentication} -- the PSU uses a different device from the one being used to request services from the AISP to authenticate and authorise the account access consent. The AISP can then receive the access token through two different approaches, callback or polling.
	\end{itemize}
	During authorisation of the consent, the PSU has to select the payment account(s) for which the chosen permissions should apply.
	\item[Step 4] As the AISP now has the access token to the account data, they have to make first a GET request to the accounts endpoint to retrieve the accessible accounts, including their unique IDs. The IDs can later be used to request the data of specific accounts.
\end{description}

The aforementioned steps authenticate the AISP, thereby granting them the right to request the ASPSP for the data the PSU has consented to. To retrieve specific PSU account data, the AISP will have to request the data via the appropriate link using the correct endpoint and method from the ASPSP.

\subsection{Account and Transaction API Use Case Example}

In Figure \ref{fig:use-case}, we present an abstract version of a user journey describing the Payment Services User (PSU) interaction with a third-party application (AISP) and the banking infrastructure (ASPSP). This use case walks through the various stages of authentication and consent provision necessary for secure data sharing between the user, application, and the bank. A concrete example within the NatWest sandbox is given in \ref{appendix:natwest-walk-through}.

Initially , the PSU (Payment Services User), representing the user, interacts with the system by launching the application (DemoAPP) that aims to access their financial data. The first interaction node is the Bank Selection screen \textcircled{1}, where the user is prompted to choose from a list of available banks. Once the user selects their bank, they proceed to the next screen.

Next, the user encounters the Consent Screen \textcircled{2}, which is of key importance for data sharing. Here, the app requests permission to access certain details from the user’s bank account. To comply with privacy policies and terms of service, the user must explicitly grant permission by clicking \textit{Continue}. If they opt to cancel the process, they can click \textit{Back}, which will abort the data-sharing attempt.

After consent is granted, the user is redirected to their bank's application \textcircled{3}, the Banking App interface provided by the ASPSP (Account Servicing Payment Service Provider). The PSU is asked to select the specific accounts they wish to share with DemoService (the AISP). The option to cancel is also available, ensuring the user maintains control over the consent process.

Once the selection is confirmed, the user is temporarily redirected via the Redirect Screen \textcircled{4}, which transitions them back to the AISP's (DemoAPP) interface. The application notifies the user that they are being redirected to complete the process.

Finally, the flow concludes at the Result Screen \textcircled{5}, where the PSU is informed that the data has been successfully retrieved from the selected account(s). The screen also provides a summary of what data was shared. 

\begin{figure}
    \centering
    
\begin{tikzpicture}[
    every node/.style={font=\footnotesize},
    psu/.style={draw, circle, fill=cyan!20, minimum size=4cm},
    app/.style={draw, line width=1.5mm, rounded corners, fill=green!20, minimum width=3.5cm, minimum height=7.5cm, text width=3.5cm, align=center},
    bank/.style={draw, line width=1.5mm, rounded corners, fill=purple!20, minimum width=3.5cm, minimum height=7.5cm, text width=3.5cm, align=center},
    redirect/.style={draw, line width=1.5mm, rounded corners, fill=blue!20, minimum width=3.5cm, minimum height=7.5cm, text width=3.5cm, align=center},
    arrow/.style={->, >=stealth},
    button/.style={draw, rounded corners, fill=gray!30, minimum width=1.5cm, minimum height=0.5cm, text centered},
]

\node[psu] (psu) {PSU (end user)};

\node[app, right=of psu] (bank_selection) {
    \textbf{{\Large\textcircled{1}}} \\[5ex]
    \textbf{DemoAPP} \\[1ex]
    \textbf{Bank Selection} \\[10ex]
    \begin{itemize}
     \setlength{\itemsep}{-2pt}
     \setlength{\parsep}{0pt}
        \item \textbf{Bank 001}
        \item Bank 002
        \item Bank 003
        \item ...
    \end{itemize}
};

\node[app, right=of bank_selection] (consent_screen) {
    \textbf{{\Large\textcircled{2}}} \\[1ex]
    \textbf{DemoAPP} \\[1ex]
    \textbf{Permission to request access} \\[1ex]
    To use this service, DemoAPP needs to access information from your accounts at Bank 001. \\[1ex]
    \textit{What we need you to share:}
    \begin{itemize}[left=0.2cm]
     \setlength{\itemsep}{-2pt}
     \setlength{\parsep}{0pt}
        \item Account Details
        \item Balance Details
        \item Account Transactions
    \end{itemize}
    \textit{Click continue to allow access to this data under our terms \& conditions and privacy policy.}
    \\[1ex]
    Back -- \textbf{Continue}
};

\node[bank, below=of consent_screen] (bank_app) {
    \textbf{{\Large\textcircled{3}}} \\[5ex]
    \textbf{ASPSP} \\[1ex]
    Select account(s) to share information with DemoService (AISP). \\[1ex]
    \begin{itemize}[left=0.2cm]
     \setlength{\itemsep}{-2pt}
     \setlength{\parsep}{0pt}
     
        \item \textbf{Current Account} \\ \textit{01-02-03 00000001}
        \item Savings Account \\ \textit{01-02-03 00000002}
        \item Credit Card Account \\ \textit{**** **** **** 1234}
    \end{itemize}
    \textit{DemoService (AISP) will access your information from your account(s) until: 01/05/2025}
        \\[1ex]
        Cancel -- \textbf{Confirm}
};

\node[redirect, left=of bank_app] (redirect) {
    \textbf{{\Large\textcircled{4}}} \\[5ex]    
    \textbf{AISP} \\[15ex]
    \textbf{Redirecting to DemoService (AISP)} \\[1ex]
    Redirecting...
};

\node[app, left=of redirect] (result_screen) {
    \textbf{{\Large\textcircled{5}}} \\[5ex]
    \textbf{DemoAPP} \\[1ex]
    \textbf{Result Screen} \\[6ex]
    \textit{Done!} \\[1ex]
    We have received the requested information from your selected account. \\[1ex]
    \textit{What we need you to share:}
    \begin{itemize}[left=0.2cm]
     \setlength{\itemsep}{-2pt}
     \setlength{\parsep}{0pt}
        \item Account Details
        \item Balance Details
        \item Account Transactions
    \end{itemize}

    \textbf{View Data} -- Home
};

\draw[arrow] (psu) -- (bank_selection);
\draw[arrow] (bank_selection) -- (consent_screen);
\draw[arrow] (consent_screen) -- (bank_app);
\draw[arrow] (bank_app) -- (redirect);
\draw[arrow] (redirect) -- (result_screen);

\node[above=0.2cm of bank_selection] (step1) {\textit{Bank Selection}};
\node[above=0.2cm of consent_screen] (step2) {\textit{Consent Screen}};
\node[below=0.2cm of bank_app] (step3) {\textit{Banking App}};
\node[below=0.2cm of redirect] (step4) {\textit{Redirect}};
\node[below=0.2cm of result_screen] (step5) {\textit{Result Screen}};

\end{tikzpicture}
    \caption{Account and Transaction API Use Case Example}
    \label{fig:use-case}
\end{figure}

\subsection{Methodology and Specification Language}\label{sub:methodology}

\begin{figure}[t]
	\resizebox{118mm}{!}{

\tikzstyle{decision} = [diamond, draw, fill=blue!10,text width=4.75em, text badly centered, node distance=1.5cm, inner sep=0pt,font=\sffamily]
\tikzstyle{block} = [rectangle, draw, fill=blue!10,text width=10em, text centered, rounded corners, minimum height=3em, minimum width=3cm,font=\sffamily]
\tikzstyle{line} = [draw, -latex',font=\sffamily]
\tikzstyle{cloud} = [draw, ellipse,fill=red!20, node distance=1.5cm, minimum height=2em, minimum width=4cm, font=\sffamily]

\begin{tikzpicture}[node distance = 1.5cm, auto]
    \node [block] (requirements) {Informal Specification and Requirements};
    \node [cloud, below of=requirements] (modelling) {Modelling};
    \node [block, below of=modelling] (anbx) {AnBx Protocol Model};
	\node [cloud, below of=anbx] (anbxc1) {AnBxC Front-End};
    \node [block, left of=anbxc1,node distance=4.75cm] (channelmode) {AnBx Channel Mode};
	\node [block, below of=anbxc1] (anb) {AnB Protocol Model};
	\node [cloud, below of=anb] (ofmc) {OFMC};
	\node [cloud, right of=anbxc1, node distance=4.75cm] (anbxc2) {AnBxC Back-End};
    \node [block, below of=anbxc2] (proverifcode) {ProVerif applied-pi Model};
	\node [cloud, below of=proverifcode] (proverif) {ProVerif};
	\node [block, fill=white!20, below of=ofmc] (1s) {single session};
	\node [block, fill=white!20, right of=1s, node distance=4.75cm] (2s) {unbounded sessions};
    \path [line,dashed] (requirements) -- (modelling);
	\path [line,dashed] (modelling) -- (anbx);
	\path [line] (channelmode) -- (anbxc1);
	\path [line] (anbx) -- (anbxc1);
	\path [line] (anbxc1) -- (anb);
    \path [line] (anb) -- node {verification}(ofmc);

    \path [line] (anbxc1) -- (anbxc2);

    \path [line] (anbxc2) -- (proverifcode);
	\path [line] (proverifcode) -- node {verification}(proverif);
	
\end{tikzpicture} 

}
	
	\caption{\label{fig:Methodology}Modelling and Verification Methodology (\protect\tikz[baseline]{\protect\draw[dashed] (0,.8ex)--++(1,0);}
		manual\quad{}\protect\tikz[baseline]{\protect\draw[] (0,.8ex)--++(1,0);}
		automatic)}
\end{figure}

Our methodology (Figure \ref{fig:Methodology}) utilises the extended Alice and Bob notation (\textit{AnBx}) \cite{DBLP:journals/istr/BugliesiCMM16} to specify an abstract model of the protocol, which is built based on the available documentation of the Account and Transaction API protocol \cite{website:OB_AIS}. This notation, which extends the \textit{AnB} language~\cite{anb}, abstracts from the implementation details like cryptographic algorithm choices, but allows the formal representation of the security-relevant properties of protocols. The intruder is modelled in the Dolev-Yao style~\cite{dolev83ieee}, and the formal semantics of \textit{AnBx} is described in~\cite{DBLP:journals/istr/BugliesiCMM16}.  

For the verification, we use the \textit{Open-Source Fixed-Point Model-Checker} (OFMC)~\cite{moedersheim2009secure}, a symbolic model-checker supporting the \textit{AnB} notation. Verification is done for a bounded number of parallel sessions. However, the \textit{AnBx} Compiler and Code Generator (AnBxC)~\cite{anbx2015} is used to pre-process the model to benefit from a stricter type system and named expression abstractions (\verb'Definitions' section within \textit{AnBx} files). 

Furthermore, the AnBx compiler can generate a typed applied-pi model that we use to verify the security goals for an unbounded number of parallel sessions with \textit{ProVerif} \cite{blanchet2001efficient}. In fact, due to the complexity of the model, OFMC is only able to verify a single session before state explosion (see Section~\ref{sec:security}). Thus, potentially missing attacks involving the attacker exploiting multiple sessions running in parallel.

It should be noted that the AnBx compiler allows for the generation of OFMC and ProVerif code from the same model. This is not only convenient, including for the cross-validation of the results, but also ensures that errors due to manual encoding of models in two different languages cannot be inadvertently introduced.

Additionally, AnBx supports strict type checking, similar to strongly-typed languages, enabling users to detect ill-formed specifications. It also conducts executability and reachability checks on the generated code which complements the checks done by the verification tools.

An \textit{AnBx} specification comprises several sections. The \verb'Types' section declares the different identifiers used in the protocol. This includes agents, constant and variable (random) numbers, and transparent functions. Transparent functions are user-defined through their signature, thereby abstracting from their implementation details (\emph{i.e.}, they are not interpreted). The \verb'Knowledge' section describes the initial data each agent has before running the protocol.

The information flow is described in the \verb'Actions' section, where details about messages exchanged by agents are specified. In general, along with the standard plain (i.e., insecure) communication channel, it is possible to specify three types of channels (also known as \emph{AnB bullet channels} as they were originally introduced in ~\cite{moedersheim2009secure}): \emph{authentic}, \emph{confidential}, and \emph{secure}, with variants that allow agents to be identified by a pseudonym rather than by a real identity. The supported channels are: 
\begin{enumerate}
	\item \verb'A -> B: M', an insecure channel from $A$ to $B$, under the
	complete control of a Dolev-Yao intruder \cite{dolev83ieee};
	\item \verb'A *-> B: M', an authentic channel from $A$ to $B$, where $B$
	can rely on the fact that $A$ has sent the message $M$ and meant
	it for $B$;
	\item \verb'A ->* B: M', a confidential channel, where $A$ can rely
	on the fact that only $B$ can receive the message $M$;
	\item \verb'A *->* B: M', a secure channel (both authentic and confidential).
\end{enumerate}

Furthermore, the model can be used to verify specific security properties (\verb'Goals' section), such as (weak and strong) authentication and secrecy goals:
\begin{itemize}[noitemsep]
	\item \verb'A weakly authenticates B on M': agent A has evidence that the message M has been endorsed by agent B with the intention to send it to A (i.e., non-injective agreement~\cite{Lowe97});
	\item \verb'A authenticates B on M': weak authentication plus evidence of the freshness of the message M (i.e., injective agreement~\cite{Lowe97});
	\item \verb'M secret between A, B': message M is kept confidential among the listed agents.
\end{itemize}

 While the Open Banking Account and Transaction API protocol describes in detail the information flow ~\cite{website:OB_AIS}, it lacks an explicit definition of the security goals that the exchanges between agents are meant to convey. Therefore, one of our tasks consisted of identifying suitable goals for the protocol model.

Presenting all the details of the modelling language is beyond the scope of this paper. However, we will introduce the relevant language features in the next section when needed to present the model.

\parahead{Vertical Protocol Composition.~} Since, in our Account and Transaction API model, the protocol runs on top of channels providing security guarantees abstraction (i.e., bullet channels, \verb'*->*'), we need to discuss whether this vertical composition is secure~\cite{Moedersheim2014}. In essence, we are composing a secure channel (namely TLS) and the Account and Transaction API protocols.

In general, given a secure protocol $P_{1}$ that provides a certain channel type as a goal and another secure protocol $P_{2}$ that assumes this channel type, their vertical composition $P_{2}[P_{1}]$ is not secure, as attacks may be possible even when the individual protocols are all secure in isolation. Sufficient conditions for vertical composition have been established by M{\"o}dersheim and Vigan{\`o}~ \cite{Moedersheim2014}, requiring the disjointness of the message formats of $P_{1}$ and $P_{2}$, and that the payloads of $P_{2}$ are embedded into $P_{1}$ under a unique context to define a sharp borderline. As detailed in \cite{Moedersheim2014}, in practice such  conditions are  satisfied  by a large class of protocols. 

As the specific implementation of the underlying protocol is not part of the Account and Transaction API Protocol ($P_{2}$), but $P_{2}$ only assumes that communication occurs on channels that guarantee secret communication with server mutual authentication ($P_{1}$), we make our analysis under the assumption that the conditions sufficient for vertical composition specified in~\cite{Moedersheim2014} are satisfied.

\section{Formal Model}
\label{sec:implementation}

We define an \textit{AnBx} model of the Account and Transaction API protocol~\cite{website:OB_AIS} to analyse and accurately verify its information flow and security goals (see~\ref{appendix:anb_model}). Our model abstracts from dependent technologies such as OAuth 2.0~\cite{report:rfc6749_auth-2.0} and OpenID Connect~\cite{Mainka2017,Navas2019}, which are specified in the Open Banking Security Profile~\cite{website:OB_security_profile_impl_draft}. This abstraction does not impact our results (Section~\ref{sec:security}) since we focus on protocol messages rather than their transport medium.  

Since confidential messages are exchanged over the Internet, the Open Banking specification~\cite{website:OB_security_profile_OIDC} mandates the use of Transport Layer Security (TLS) between all parties, similar to the Financial API specification~\cite{report:fapi_security_profile_part1}. Rather than verifying TLS, which is beyond our scope, we assume its security guarantees and model confidential and authenticated communication using bullet channels (\verb'*->*'). These serve as an abstraction for a secure, mutually authenticated TLS channel~\cite{moedersheim2009secure}, aligning with the vertical protocol composition discussed above. Furthermore, OAuth 2.0 and OpenID Connect have already been extensively analysed~\cite{fett2019extensive}.

It should be noted that for the current version of the Account and Transaction API Specification, requests and responses are not necessarily be digitally signed. The documentation~\cite{website:R/W_Data} explicitly states that this is optional. To achieve non-repudiation, maintaining digital records and evidence would be challenging if the API relied solely on TLS. Instead, there is a provision for using a different mechanism to maintaining digital records, such as JSON Web Signature (JWS), defined in RFC 7515 -- Appendix F \cite{rfc7515}. Additionally, implementers of the standards can optionally add signatures and further message encryption, but the applicability to individual requests and responses is not defined in the standard itself.  Therefore, we encode the baseline model without the addition of an explicit layer of digital signature nor optional cryptographic features. Moreover, our model specifically considers the simpler Authorisation Code Grant variant in Step 3, as the Client-Initiated back channel Authentication variant relies on communications done on another device.

In other words, we focus our analysis on the essence of AISP message exchanges with respect to identified security goals. We provide a detailed explanation of our formal model in \textit{AnBx} next. A sequence diagram containing action labels is given in Figure~\ref{fig:anb_aisp_seq_diagram}.

\begin{figure}[t]
\centering

	\tikzset{
	every picture/.append style={
		transform shape,
		scale=0.62
	}
}

\begin{sequencediagram}
	
	\renewcommand\unitfactor{0.48}
	
	\newinst{PSU}{PSU}
	\newinst[3]{AISP}{AISP}
	\newinst[3]{ASPSPAuth}{ASPSP Auth Server}
	\newinst[3]{ASPSPResource}{ASPSP Resource Server}
	
	\begin{sdblock}{Step 1}{Request account information}
		\mess{PSU}{A.1.1 IntentAgreement,PSU,aspspA }{AISP}
	\end{sdblock}
			
	\begin{sdblock}{Step 2}{Setup account request}		
		\mess{AISP}{A2.1 ClientTokenReq}{ASPSPAuth}
		\mess{ASPSPAuth}{A2.2 ClientTokenRes}{AISP}
		\mess{AISP}{A2.3 IntentReq}{ASPSPResource}
		\mess{ASPSPResource}{A2.4 Intent}{AISP}
	\end{sdblock}
		
	\begin{sdblock}{Step 3}{Authorise consent (Authorisation Code Grant)}
		\begin{sdblock}{Step 3.1}{Request initiation of consent authorisation}
			\mess{AISP}{A3.1.1 RedirectToASPSPCmd}{PSU}
			\mess{PSU}{A3.1.2 InitAuthIntentReq}{ASPSPAuth}
		\end{sdblock}
		\begin{sdblock}{Step 3.2}{Review and authorise consent resource}
			\mess{ASPSPAuth}{A3.2.1 RetriveIntentReq}{ASPSPResource}
			\mess{ASPSPResource}{A3.2.2 RetriveIntentRes}{ASPSPAuth}
			\mess{ASPSPAuth}{A3.2.3 RetriveIntentRes}{PSU}
			\mess{PSU}{A3.2.4 SelectedAccounts}{ASPSPAuth}
			\mess{ASPSPAuth}{A3.2.5 AuthoriseIntentReq}{ASPSPResource}
			\mess{ASPSPResource}{A3.2.6 AuthoriseIntentRes}{ASPSPAuth}
		\end{sdblock}
		\begin{sdblock}{Step 3.3}{Request authorisation token}
			\mess{ASPSPAuth}{A3.3.1 RedirectToAISPCmd}{PSU}
			\mess{PSU}{A3.3.2 RedirectToAISPCmd}{AISP}
			\mess{AISP}{A3.3.3 AuthTokenReq}{ASPSPAuth}
			\mess{ASPSPAuth}{A3.3.4 AuthToken}{AISP}
		\end{sdblock}	
	\end{sdblock}
	
	\begin{sdblock}{Step 4}{Request data}
		\mess{AISP}{A4.1 AccountsReq}{ASPSPResource}
		\mess{ASPSPResource}{A4.2 Accounts,PSU}{AISP}
	\end{sdblock}

\end{sequencediagram}

\caption{Account and Transaction API Protocol -- Model Sequence Diagram}
\label{fig:anb_aisp_seq_diagram}
\end{figure}
\subsection{Roles and Responsibilities}\label{ssec:roles-and-respo}

The protocol has multiple participating agents. Upon interaction with an application developed by an AISP, the PSU initiates the protocol aiming to grant the AISP limited access to their account data. The AISP objective is to obtain access to the PSU's account data to provide the user with a service. The ASPSP has two separate roles: the authorisation server (\verb'aspspA'), which authenticates AISPs and PSUs to generate tokens enabling AISPs to access endpoints; and the resource server (\verb'aspspR'), which maintains resources, such as consents (i.e., authorisation message) and PSU account data. 

By observing the responsibilities of each agent, the authorisation and resource servers must be trusted parties: if any of them acts maliciously, the protocol can be trivially broken. Moreover, although the authorisation and resource servers are typically run by the same institution (e.g., a bank), we assume that in practice the authorisation and resource servers will run independently and will use different cryptographic identities. In fact, it would be a bad engineering practice to reuse the same cryptographic keys for different purposes \cite{abadi1994pep}. In our model, we define these roles as follows: 

\begin{lstlisting}[language=AnBx]
Agent PSU, AISP, aspspA, aspspR;
\end{lstlisting}

It should be noted that trusted agents, i.e., roles that cannot be impersonated by the attacker, are represented  in \textit{AnBx}  by identifiers beginning with a lowercase letter. That is, we are saying that the authorisation (\verb'aspspA') and resource (\verb'aspspR') servers are trust and distinct entities.

\subsection{Initial Knowledge}
The PSU and AISP know each other given that, since the PSU has been in contact with the AISP to exchange permissions and to inform the search provider or financial institution of the ASPSP of interest to be involved in the transaction (see Bank Selection~\textcircled{1} in Figure~\ref{fig:use-case}). The AISP is assumed to have identified the ASPSP's corresponding authorisation server \verb'aspspA'.
The authorisation and resource servers are known to each other and their identities are also known by the PSU, as the ASPSP is the entity where the PSU's account data is stored.

\begin{lstlisting}[language=AnBx]
PSU   : PSU, AISP, aspspA, aspspR;
AISP  : AISP, PSU, aspspA, aspspR;
aspspA: PSU, AISP, aspspA, aspspR, fClientCredToken, fAuthCode,      fAuthCodeToken;
aspspR: PSU, AISP, aspspA, aspspR, fCreateIntent, fGetIntent,     fFetchAccounts, fAuthoriseIntent,fAccountsEndpoint;

aspspA,PSU  share fPSUSecret(PSU,aspspA);
aspspA,AISP share fAISPSecret(aspspA,AISP);
\end{lstlisting}

We provide here some key information about the initial knowledge, giving additional details about the declared functions when we explain the protocol run. %

\parahead{PSU/AISP}~Authentication requires the PSU and AISP to pre-share a secret with the authorisation server \verb'aspspA'. This secret is known a priori as represented by transparent private function calls
 \verb'fPSUSecret(PSU,aspspA)' and \verb'fAISPSecret(AISP,aspspA)', which allows us to abstract from the secret agreement mechanism, which in its simplest form that could be a password. We use two distinct functions, as, in general, the authentication systems for account holders (PSU) and third-party applications may be different. The symbol \verb'->*' denotes private functions in \textit{AnBx}, functions that are not available to the intruder.

\begin{lstlisting}[language=AnBx]
	Function [Agent,Agent ->* Number] fPSUSecret;
	Function [Agent,Agent ->* Number] fAISPSecret;
\end{lstlisting}

\parahead{Authorisation and Resource Servers}~The authorisation server performs numerous operations over the protocol execution through transparent functions:
~\verb'fClientCredToken' generates a token acquired through a client credential grant;~ \verb'fAuthCode' generates an authorisation code;~and \verb'fAuthCodeToken' generates a token acquired through the authorisation code. At each protocol run, these functions return a different value, as they are parameterised on data that changes at each protocol run. Definition \verb'IntentAgreement' contains the information agreed by PSU and AISP. This is modelled as a random value that is used to model the OAuth token issued by the authorisation server. This will be next used by function \verb'fClientCredToken' and definition~\verb'AISPSecretNonce', detailed later. Lack of such randomness was a key weakness of our previous work~\cite{abz2020}. 

The authorisation server is also aware of its PSU and AISP credentials to authenticate them. The resource server performs state-changing operations through abstract functions:~\verb'fCreateIntent' and \verb'fGetIntent' to create and retrieve a consent resource, respectively;~\verb'fFetchAccounts' retrieves a list of all PSU accounts;~\verb'fAuthoriseIntent' updates consent authorisation;~and \verb'fAccountsEndpoint' returns the PSU account data.

\parahead{Roles Restriction}~To make our model realistic, we impose restrictions on the role performed by different agents. To declare which agent is not allowed to act as another agent, we can use the~\verb'where' keyword at the end of the \emph{Knowledge} section.

\begin{lstlisting}[language=AnBx]
   where PSU!=AISP, PSU!=aspspA,PSU!=aspspR, AISP!=aspspA, AISP!=aspspR
\end{lstlisting}

For example, the PSU cannot act as the AISP, which is unrealistic since, in the UK, AISPs are regulated by the Financial Conduct Authority (FCA)~\cite{website:FCA_about}.  
We also excluded the possibility of the PSU and the AISP acting as either the authorisation or resource servers, as these servers are considered trusted and must be distinct. Although it is not necessary to explicitly enforce these restrictions in the model, in \textit{AnBx}, trusted agents cannot be impersonated by other agents or the intruder by default. However, we observed that verification with OFMC is faster when these conditions are explicitly stated. For example, simply removing the condition \verb'AISP!=aspspR' increased the verification time by more than 90 times.

\subsection{Actions}

\begin{table}[th!]
	\centering
	\footnotesize
	\begin{tabular}{|>{\arraybackslash}m{0.9cm}|>{\arraybackslash}m{2.05cm}|>{\raggedright\arraybackslash}m{7.9cm}|}
		\hline
		\textit{Action} & \textit{Agent} & \textit{Message Description} \\
		\hline
		A1.1 & PSU & \verb'IntentAgreement' represents the agreement between AISP and PSU. \\
		\hline
		A2.1 & AISP & \verb'ClientTokenReq' represents the request for a client token. \\
		\hline
		A2.2 & Auth. Server & \verb'ClientTokenRes' represents the response to a valid client token request. \\
		\hline
		A2.3 & AISP & \verb'IntentReq' represents the request to create a consent resource. \\
		\hline
		A2.4 & Resource Server & \verb'Intent' represents the response to a valid consent creation request. \\
		\hline
		A3.1.1 & AISP & \verb'RedirectToASPSPCmd' serves as the command to redirect the PSU to the authorisation server. \\
		\hline
		A3.1.2 & PSU & \verb'InitAuthIntentReq' represents the request to initiate consent authorisation. \\
		\hline
		A3.2.1 & Auth. Server & \verb'RetrieveIntentReq' represents the request to retrieve the consent resource and PSU accounts. \\
		\hline
		A3.2.2 & Resource Server & \verb'RetrieveIntentRes' represents the response to a request to retrieve the consent resource and PSU accounts. \\
		\hline
		A3.2.3 & Auth. Server & \verb'RetrieveIntentRes' is defined by the resource server in the previous step. \\
		\hline
		A3.2.4 & PSU & \verb'SelectedAccounts' is the set of accounts the PSU has selected to associate with consent. \\
		\hline
		A3.2.5 & Auth. Server & \verb'AuthoriseIntentReq' serves as the request to update the state of the consent resource to authorised. \\
		\hline
		A3.2.6 & Resource Server & \verb'AuthoriseIntentRes' serves as the response to a consent resource-authorisation request. \\
		\hline
		A3.3.1 & Auth. Server & \verb'RedirectToAISPCmd' represents the command to redirect the PSU back to the AISP. \\
		\hline
		A3.3.2 & PSU & \verb'RedirectToAISPCmd' is defined by the authorisation server in the previous step. \\
		\hline
		A3.3.3 & AISP & \verb'AuthTokenReq' represents the request for an authorisation token. \\
		\hline
		A3.3.4 & Auth. Server & \verb'AuthToken' serves as the response to a valid authorisation token request. \\
		\hline
		A4.1 & AISP & \verb'AccountsReq' represents the request for general PSU account data. \\
		\hline
		A4.2 & Resource Server & \verb'AccountsRes' represents the response to a valid PSU account data retrieval request. \\
		\hline
	\end{tabular}
	\caption{Mapping of Actions to Agents and Messages}
	\label{tab:action_agent_mapping}
\end{table}

The protocol actions are described next, as illustrated in Figure~\ref{fig:anb_aisp_seq_diagram}. All exchanges are assumed to be executed over a secure channel (i.e. \verb|*->*|), and actions are labelled according to each protocol step (see Table~\ref{tab:action_agent_mapping}). The protocol consists of four steps: 1)~\emph{Account Information Request}; 2)~\emph{Account Access Consent Setup}; 3)~\emph{Consent Authorisation}; and 4)~\emph{Data Request}. The complete protocol source can be found in~\ref{appendix:anb_model}.  

\begin{figure}[th!]
\centering

	\tikzset{
	every picture/.append style={
		transform shape,
		scale=0.62
	}
}
\vspace{4mm}
\begin{sequencediagram}
	
	\renewcommand\unitfactor{0.48}
	
	\newinst{PSU}{PSU}
	\newinst[3]{AISP}{AISP}
	\newinst[3]{ASPSPAuth}{ASPSP Auth Server}
	\newinst[3]{ASPSPResource}{ASPSP Resource Server}
	
	\begin{sdblock}{Step 1}{Request account information}
		\mess{PSU}{A.1.1 IntentAgreement,PSU,aspspA }{AISP}
	\end{sdblock}
			
\end{sequencediagram}

\caption{Account and Transaction API Protocol -- Model Sequence Diagram -- Step 1}
\label{fig:anb_aisp_seq_diagram-step1}
\end{figure}

\parahead{Step 1: Account Information Request}~(Figure \ref{fig:anb_aisp_seq_diagram-step1})~First, the PSU informs the AISP of them wanting to share account information through an intent agreement (\textit{A1.1}), which is exchanged along with the PSU and the authorisation server identities. This information is assumed to be the result of an interaction, prior to the protocol run, between AISP and PSU. In particular, AISPs must ask the PSU to identify their ASPSP before requesting consent, so that the consent request can be constructed in line with the ASPSP’s data capabilities. The AISP must provide the PSU with a description of the data being requested and the permissions that need to be granted (e.g. reading account details, regular payments, transactions, statements, etc.).
The PSU then chooses which permissions to accept and adds further restrictions on data access if required, thereby obtaining the intent agreement. Once the PSU has consented, the PSU will be directed to their ASPSP. As the exact information included in the intent agreement varies depending on the PSU choices, we have abstracted here the intent as a nonce of type \verb|IntentNo|, as it will, in any case, include data to distinguish one request from another through random values.

\begin{figure}[h!]
\centering

	\tikzset{
	every picture/.append style={
		transform shape,
		scale=0.62
	}
}

\begin{sequencediagram}
	
	\renewcommand\unitfactor{0.48}
	
	\newinst{PSU}{PSU}
	\newinst[3]{AISP}{AISP}
	\newinst[3]{ASPSPAuth}{ASPSP Auth Server}
	\newinst[3]{ASPSPResource}{ASPSP Resource Server}

	\begin{sdblock}{Step 2}{Setup account request}		
		\mess{AISP}{A2.1 ClientTokenReq}{ASPSPAuth}
		\mess{ASPSPAuth}{A2.2 ClientTokenRes}{AISP}
		\mess{AISP}{A2.3 IntentReq}{ASPSPResource}
		\mess{ASPSPResource}{A2.4 Intent}{AISP}
	\end{sdblock}

\end{sequencediagram}

\caption{Account and Transaction API Protocol -- Model Sequence Diagram -- Step 2}
\label{fig:anb_aisp_seq_diagram-step2}
\end{figure}

\parahead{Step 2: Account Access Consent Setup}~(Figure \ref{fig:anb_aisp_seq_diagram-step2})~To be able to create a consent resource, the AISP informs the authorisation server that it requires an access token and provides the necessary data for such a token (\textit{A2.1}). The authorisation server provides the AISP with the access token, here referred to as the client token, to request the creation of a consent resource from the resource server (\textit{A2.2}). This is done after the authorisation server internally authenticates the AISP on their credentials. After that, the AISP asks the resource server to create a consent resource based on the agreement between the AISP and PSU (\textit{A2.3}). This consent setup relies on the correctness of OAuth 2.0. Namely, the freshness and authenticity of various tokens involved.

The requirements lack information regarding the communication between the authorisation and resource servers. As a result, we assume that these servers are responsible for their roles only and that any limited and controlled access to each other's resources, if it exists, does not have any side effect on the protocol run.
In this case, when the authorisation server requires access to a consent or the PSU accounts, it is assumed that it has to request such information from the resource server. After creating a consent resource based on the request of the AISP, the resource server provides the resource including its identifier to the AISP (\textit{A2.4}) for later reference.

\begin{figure}[t]
\centering

	\tikzset{
	every picture/.append style={
		transform shape,
		scale=0.62
	}
}

\begin{sequencediagram}
	
	\renewcommand\unitfactor{0.48}
	
	\newinst{PSU}{PSU}
	\newinst[3]{AISP}{AISP}
	\newinst[3]{ASPSPAuth}{ASPSP Auth Server}
	\newinst[3]{ASPSPResource}{ASPSP Resource Server}

	\begin{sdblock}{Step 3}{Authorise consent (Authorisation Code Grant)}
		\begin{sdblock}{Step 3.1}{Request initiation of consent authorisation}
			\mess{AISP}{A3.1.1 RedirectToASPSPCmd}{PSU}
			\mess{PSU}{A3.1.2 InitAuthIntentReq}{ASPSPAuth}
		\end{sdblock}
		\begin{sdblock}{Step 3.2}{Review and authorise consent resource}
			\mess{ASPSPAuth}{A3.2.1 RetriveIntentReq}{ASPSPResource}
			\mess{ASPSPResource}{A3.2.2 RetriveIntentRes}{ASPSPAuth}
			\mess{ASPSPAuth}{A3.2.3 RetriveIntentRes}{PSU}
			\mess{PSU}{A3.2.4 SelectedAccounts}{ASPSPAuth}
			\mess{ASPSPAuth}{A3.2.5 AuthoriseIntentReq}{ASPSPResource}
			\mess{ASPSPResource}{A3.2.6 AuthoriseIntentRes}{ASPSPAuth}
		\end{sdblock}
		\begin{sdblock}{Step 3.3}{Request authorisation token}
			\mess{ASPSPAuth}{A3.3.1 RedirectToAISPCmd}{PSU}
			\mess{PSU}{A3.3.2 RedirectToAISPCmd}{AISP}
			\mess{AISP}{A3.3.3 AuthTokenReq}{ASPSPAuth}
			\mess{ASPSPAuth}{A3.3.4 AuthToken}{AISP}
		\end{sdblock}	
	\end{sdblock}

\end{sequencediagram}

\caption{Account and Transaction API Protocol -- Model Sequence Diagram -- Step 3}
\label{fig:anb_aisp_seq_diagram-step3}
\end{figure}

\parahead{Step 3: Consent authorisation}~(Figure \ref{fig:anb_aisp_seq_diagram-step3})~Authorisation is defined in $3$ stages:~1)~initiation of consent authorisation;~2)~review and authorise the consent;~and 3)~obtain an access token to the PSU's account data.

After creating a consent resource, the AISP instructs the PSU to redirect to the authorisation server with the required data to request consent authorisation (\textit{A3.1.1}). The PSU then redirects to the authorisation server and requests the consent authorisation (\textit{A3.1.2}).

After internally authenticating the PSU, the authorisation-server requests from the resource-server resources needed to be reviewed by the PSU before authorising the consent (\textit{A3.2.1}). The resource server provides the authorisation server with the requested resources (\textit{A3.2.2}). The authorisation-server then forwards these resources to the PSU to review and authorise the consent (\textit{A3.2.3}), and inform the authorisation-server of which of their accounts to associate with the consent (\textit{A3.2.4}).

After the PSU authorises the consent, the authorisation server forwards the authorisation process along with the required data to the resource server (\textit{A3.2.5}). This is because the authorisation server does not update the consent resources, as it is not its responsibility to maintain resources, as previously assumed. Furthermore, sub-step $40$ of the requirements sequence diagram in~\cite{website:OB_security_profile_impl_draft} (and also in Figure~\ref{fig:anb_aisp_seq_diagram}), shows the authorisation server explicitly requesting the resource server to update the consent resource. After authorising the consent resource, the resource server updates the consent state and informs the authorisation server of their success (\textit{A3.2.6}).

When informed that the consent resource has been authorised, the authorisation server instructs the PSU to redirect to the AISP with the required data to request an access token (\textit{A3.3.1}). The PSU redirects back to the AISP, providing it with the information required to obtain an access token to the PSU account data, referred to as an authorisation token (\textit{A3.3.2}). The AISP then requests such a token from the authorisation server (\textit{A3.3.3}):~part of the data used for the request is only obtained after consent authorisation. Subsequently, the authorisation server provides the AISP with the access token (\textit{A3.3.4}). This is done after the authorisation server internally authenticates the AISP on their credentials.

\begin{figure}[t]
\centering

	\tikzset{
	every picture/.append style={
		transform shape,
		scale=0.62
	}
}

\begin{sequencediagram}
	
	\renewcommand\unitfactor{0.48}
	
	\newinst{PSU}{PSU}
	\newinst[3]{AISP}{AISP}
	\newinst[3]{ASPSPAuth}{ASPSP Auth Server}
	\newinst[3]{ASPSPResource}{ASPSP Resource Server}
	
	\begin{sdblock}{Step 4}{Request data}
		\mess{AISP}{A4.1 AccountsReq}{ASPSPResource}
		\mess{ASPSPResource}{A4.2 Accounts,PSU}{AISP}
	\end{sdblock}

\end{sequencediagram}

\caption{Account and Transaction API Protocol -- Model Sequence Diagram -- Step 4}
\label{fig:anb_aisp_seq_diagram-step4}
\end{figure}

\parahead{Step 4: Data request}~(Figure \ref{fig:anb_aisp_seq_diagram-step4})~As the AISP now has the access token, they request the resource server to return the permitted PSU account data (\textit{A4.1}). The resource server then obtains and returns such data to the AISP (\textit{A4.2}).

\subsection{Messages Definitions}
\label{sub:messages_definitions}

\begin{table}[t]
	\centering
	\footnotesize
	\begin{tabular}{|>{\arraybackslash}m{1.70cm}|l|l|>{\raggedright\arraybackslash}m{4.8cm}|}
		\hline
		\textit{Agent}  & \textit{Action} & \textit{Message ID} & \textit{Definition Expansion} \\
		\hline
		PSU  & A1.1 & \verb'IntentAgreement' & -- \\
		\hline
		PSU  & A3.1.2 & \verb'PSUCredentials' & PSU, fPSUSecret(PSU, aspspA) \\
		\hline
		PSU  & A3.1.2 & \verb'InitAuthIntentReq' & RedirectToASPSPCmd, PSUCredentials \\
		\hline
		PSU  & A3.2.4 & \verb'SelectedAccounts' & IntentAgreement \\
		\hline
		AISP  & A2.1 & \verb'ClientTokenReq' & AISPCredentials \\
		\hline
		AISP  & A2.1 & \verb'AISPCredentials' & AISP, fAISPSecret(AISP, aspspA) \\
		\hline
		AISP  & A2.3 & \verb'IntentReq' & ClientToken, IntentAgreement \\
		\hline
		AISP  & A3.1.1 & \verb'RedirectToASPSPCmd' & ASPSPAuthPSUEndpoint, AISPEndpoint, Intent \\
		\hline
		AISP  & A3.3.3 & \verb'AuthTokenReq' & AISPCredentials, AuthorisationCode \\
		\hline
		AISP  & A4.1 & \verb'AccountsReq' & AuthToken \\
		\hline
		Auth. Server  & A2.2 & \verb'ClientTokenRes' & AISPSecretNonce,ClientToken \\
		\hline
		Auth. Server  & A2.2 & \verb'ClientToken' & fClientCredToken(AISP, AISPSecretNonce) \\
		\hline
		Auth. Server  & A3.2.1 & \verb'RetrieveIntentReq' & Intent, PSU \\
		\hline
		Auth. Server  & A3.3.1 & \verb'RedirectToAISPCmd' & AISPEndpoint, AuthorisationCode \\
		\hline
		Auth. Server  & A3.3.1 & \verb'AuthorisationCode' & fAuthCode(Intent) \\
		\hline
		Auth. Server  & A3.3.4 & \verb'AuthToken' & fAuthCodeToken(AuthorisationCode) \\
		\hline
		Res. Server  & A2.4 & \verb'Intent' & fCreateIntent(ClientToken, IntentAgreement) \\
		\hline
		Res. Server  & A3.2.2 & \verb'RetrieveIntentRes' & fGetIntent(Intent), fFetchAccounts(PSU) \\
		\hline
		Res. Server  & A3.2.6 & \verb'AuthoriseIntentRes' & fAuthoriseIntent(PSU, Intent, SelectedAccounts) \\
		\hline
		Res. Server  & A4.2 & \verb'Accounts' & fAccountsEndpoint(AuthToken, AccountsSecretNonce, empty) \\
		\hline
	\end{tabular}
	\caption{Message Definitions by Originating Agent and Corresponding Action}
	\label{tab:message_definitions}
\end{table}

We explain here the structure of each message exchanged (Table \ref{tab:message_definitions}) and discussed in the actions subsection. It covers the definitions each agent relies on to form their messages. Furthermore, this section also breaks down the types used by the agents to form \textit{AnBx} definitions. To link the messages with the actions throughout this subsection, we rely on the labelling of actions (e.g. \textit{A2.1}), as per the \textit{AnBx} model related to the sequence diagram in Figure \ref{fig:anb_aisp_seq_diagram}.

In this model, we abstract away from data exchanged between the agents that do not affect the model verification process. For example, HTTP response codes and data type header fields are ignored.

\parahead{PSU} \textit{Action A1.1} --~\verb'IntentAgreement' represents the complete agreement between AISP and PSU, assumed to be sent by the PSU. This assumes that the PSU already has the list of permissions asked by the AISP. The PSU then chooses which permissions to accept and adds additional restrictions on data access if required, thereby obtaining the~\verb'IntentAgreement'. We assume this agreement could differ between different protocol runs. Note that the intent agreement process between the AISP and the PSU is not defined by the protocol specification.

There are two options to generate ~\verb'IntentAgreement', either by defining it as a random independent \verb'Number' variable or a \verb'Number' variable dependent on some parameters passed to a function. The mechanism for which permissions a PSU accepts is unspecified in the standard and depends on the specific way the TTPs developer designs the AISP/PSU interaction. Therefore, an independent variable is used to represent ~\verb'IntentAgreement'. %

\begin{lstlisting}[language=AnBx]
    IntentNo IntentAgreement;
\end{lstlisting}

\noindent\textit{Action A3.1.2} --~\verb'InitAuthIntentReq' represents the request to initiate consent authorisation. It contains the redirection data defined and sent by the AISP, and the PSU's credentials required to authenticate the PSU. The PSU sends both as one message to initiate consent authorisation.

\begin{lstlisting}[language=AnBx]
    InitAuthIntentReq : RedirectToASPSPCmd,PSUCredentials;
    PSUCredentials    : PSU,fPSUSecret(PSU,aspspA);
    RedirectToASPSPCmd: ASPSPAuthPSUEndpoint,AISPEndpoint,Intent;
\end{lstlisting}

~\verb'PSUCredential' represents the shared credentials when the PSU registered with the ASPSP. The~\verb'fPSUSecret' abstract function returns the secret of the credentials agreed on between the authorisation server and the PSU whose identifiers are passed into the function. Note that this is an abstract (private) function (denoted as \verb'->*') and therefore not available to any agents or the intruder. Only the PSU and the ASPSP know that value (i.e., akin to a shared password).\label{fSecrets_def}
\noindent ~\verb'RedirectToASPSPCmd' is defined by the AISP and is a command that tells the PSU to redirect to the ASPSP end-point, and includes the \verb'Intent' computed in the previous steps (and explained later). It should be noted that ~\verb'Intent' is computed based on the ~\verb'IntentAgreement' which is different at every run of the protocol.

\begin{lstlisting}[language=AnBx]
    Function [Agent,Agent ->* Number] fPSUSecret;
    Function [Agent,Agent ->* Number] fAISPSecret;
\end{lstlisting}

\noindent The protocol's sequence diagram (Figure \ref{fig:anb_aisp_seq_diagram}) shows that the PSU is redirected to the authorisation server after which they provide their credentials to be authenticated. However, we abstract from this as explained for~\verb'InitAuthIntentReq', where the PSU sends the redirection data and their credentials as one message. %

\vspace{5pt}

\noindent \textit{Action A3.2.4} --~\verb'SelectedAccounts' is the set of accounts that the PSU has selected to associate with consent. The selection is from the list of all PSU accounts forwarded by the authorisation server to the PSU. For practical reasons, we assume the PSU select all accounts in the~\verb'IntentAgreement'. In reality, the (human) user can select a subset of the presented accounts (see screen \textcircled{3} in Figure \ref{fig:use-case}).

\begin{lstlisting}[language=AnBx]
    SelectedAccount: IntentAgreement;
\end{lstlisting}

\noindent \textit{Action A3.3.2} -- ~\verb'RedirectToAISPCmd' is defined by the authorisation server as already explained for \textit{Action A3.1.2}.%

\parahead{AISP} \textit{Action 2.1} -- ~\verb'ClientTokenReq' represents the request for a client token. It includes the AISP's credentials to be authenticated by the authorisation server. The credentials are necessary because it is a requirement of the specification to authenticate the AISP at the token endpoint \cite{website:OB_security_profile_OIDC}. As the token requested is acquired through a client credential grant, we refer to it as a client token.

\begin{lstlisting}[language=AnBx]
    AISPCredentials: AISP, fAISPSecret(AISP,aspspA);
    ClientTokenReq : AISPCredentials;
\end{lstlisting}

\noindent The specification offers three ways to authenticate the AISP at a token endpoint~\cite{website:OB_security_profile_OIDC}. We model the weakest allowed approach, which is authenticating based on an exchanged client secret (method 2). Choosing the authentication approach affects how we perceive the data the AISP exchanges with the authorisation server to be authenticated.

~\verb'AISPCredentials' represents the credentials agreed upon when the AISP registered with the ASPSP. The function~\verb'fAISPSecret' is abstract and private and returns the secret of the credentials agreed between the authorisation server and the AISP whose identifier is passed into the function. %

\begin{lstlisting}[language=AnBx]
    Function [Agent,Agent ->* Number] fAISPSecret;
\end{lstlisting}

\noindent\textit{Action A2.3} - ~\verb'IntentReq' represents the request to create a consent resource. It contains the details of the agreement passed by the PSU and an access token. As explained for the Account and Transaction API flow in Section \ref{sub:atp}, an AISP requires an access token acquired through a client credential grant to request a consent to be created. The access token is the client token acquired previously from the authorisation server.

\begin{lstlisting}[language=AnBx]
    IntentReq: ClientToken,IntentAgreement;
\end{lstlisting}

\noindent\textit{Action A3.1.1} -- ~\verb'RedirectToASPSPCmd' serves as the command to redirect the PSU to the authorisation server. It is defined as three components: the endpoints of the authorisation and AISP servers and the consent resource to authorise.

\begin{lstlisting}[language=AnBx]
    RedirectToASPSPCmd: ASPSPAuthPSUEndpoint,AISPEndpoint,Intent;
\end{lstlisting}

\noindent We include the endpoints here because the PSU relies on them to be redirected to the authorisation server and back to the AISP (see screen \textcircled{4} in Figure \ref{fig:use-case}). If those endpoints can be maliciously modified, the PSU would contact the wrong server. This is useful for the goals of the model. The endpoints are modelled as random numbers of type~\verb'URI'.

\begin{lstlisting}[language=AnBx]
    URI ASPSPAuthPSUEndpoint,AISPEndpoint;
\end{lstlisting}

\noindent As for~\verb'Intent' in~\verb'RedirectToASPSPCmd', it originally represents the consent resource created sent back to the AISP by the resource server. We abstract and use it here as the identifier of the consent. This identifier is needed by the authorisation server to determine which consent to authorise.

\vspace{5pt}

\noindent \textit{Action A3.3.3} -- ~\verb'AuthTokenReq' represents the request for an authorisation token. It includes the AISP's credentials to be authenticated and the authorisation code generated and sent by the authorisation server indirectly through the PSU. Similarly to the client token request, as the AISP is contacting a token endpoint, it is required that they are authenticated. As the token being requested is acquired through an authorisation code grant, we refer to it as an authorisation token. %

\begin{lstlisting}[language=AnBx]
    AuthTokenReq: AISPCredentials,AuthorisationCode;
\end{lstlisting}

\noindent\textit{Action A4.1} --~\verb'AccountsReq' represents the request for general PSU account data. It requires and thus includes the authorisation token generated and sent by the authorisation server. It should be noted that in our model, we cannot select individual accounts, so we assume the request is for all accounts. This explains why the definition of \verb'AccountsReq' includes only the authorisation token \verb'AuthToken'.

\begin{lstlisting}[language=AnBx]
    AccountsReq: AuthToken;
\end{lstlisting}

\parahead{Authorisation Server} In Section 6 of \cite{website:OB_security_profile_OIDC}, the specification mandates that the ASPSP verifies ownership of tokens when used by AISPs. This is an additional security layer in case an access token is compromised. For an ASPSP to be able to verify such a condition, the ASPSP must maintain a mapping between AISPs and their issued tokens. It is important to note that we have two types of access tokens and thus require multiple mappings.

Furthermore, referring to the OAuth 2.0 Authorization Framework document, it is stated that the authorisation server must ensure that an authorisation code provided by an AISP was issued to that AISP \cite[P.29]{report:rfc6749_auth-2.0}. As a result, the ASPSP also needs to keep track of who an authorisation code was issued to, presumably using a mapping. Additionally, when the authorisation server receives an authorisation code, it needs to determine the consent associated with the code to generate an access token bound to said consent. This is shown in substep 39 of the Account API Specification sequence diagram \cite{website:OB_security_profile_impl_draft}.\label{auth_code_link_def}

\vspace{5pt}

\noindent \textit{Action A2.2} -~\verb'ClientTokenRes' represents the response to a valid client token request. The definitions includes \verb'ClientToken', which is a client token generated by the authorisation server.

\begin{lstlisting}[language=AnBx]
    AISPCredentials: AISP, fAISPSecret(aspspA,AISP);
    ClientToken    : fClientCredToken(AISP,AISPSecretNonce);
    ClientTokenRes : AISPSecretNonce,ClientToken;
\end{lstlisting}

\noindent To generate the client token, the authorisation server relies on a function represented in our model by~\verb'fClientCredToken'. This function takes in the AISP's client-id and the~\verb'AISPSecretNonce' to link generated client tokens to the AISPs to which they were issued. This enables the resource server to verify the ownership of the client tokens, which is a requirement as previously stated.

\begin{lstlisting}[language=AnBx]
    Function [Agent,UUID -> JWT] fClientCredToken;
\end{lstlisting}

\noindent\textit{Action A3.2.1} -- ~\verb'RetrieveIntentReq' represents the request to retrieve the consent resource to authorise and the list of PSU accounts. It contains the identifiers of the consent resource and the PSU, to allow the resource server to find the resources.

\begin{lstlisting}[language=AnBx]
    RetrieveIntentReq: Intent,PSU;
\end{lstlisting}

\noindent\textit{Action A3.2.3} -- ~\verb'RetrieveIntentRes' is defined by the resource server (see \textit{Action A3.2.2}).

\vspace{5pt}

\noindent\textit{Action A3.2.5} -- ~\verb'AuthoriseIntentReq' serves as the request to update the state of the consent resource to authorised. It includes the PSU's identifier, consent's identifier, and PSU's selected accounts to bind to the consent.

\begin{lstlisting}[language=AnBx]
    AuthoriseIntentReq: PSU,Intent,SelectedAccounts;
\end{lstlisting}

\noindent\textit{Action A3.3.1} -- ~\verb'RedirectToAISPCmd' represents the command to redirect the PSU back to the AISP. It contains ~\verb'AISPEndpoint', which is the AISP's endpoint to redirect to, which was sent indirectly by the AISP. It also includes ~\verb'AuthorisationCode', which is an authorisation code generated by the authorisation server. The AISP requires the code to request an authorisation token.

\begin{lstlisting}[language=AnBx]
    AuthorisationCode: fAuthCode(Intent);
    RedirectToAISPCmd: AISPEndpoint,AuthorisationCode;
\end{lstlisting}
\noindent To generate the authorisation code, the authorisation server uses the ~\verb'fAuthCode' function. The function takes in the identifier of the authorised consent to link the code to. \verb'Intent' depends on \verb'IntentAgreement'. Therefore, it carries information about the selected accounts, as we assume that all listed accounts are selected.

\begin{lstlisting}[language=AnBx]
    Function [JWT -> JWT] fAuthCode;
\end{lstlisting}

\noindent\textit{Action A3.3.4} --~\verb'AuthToken' serves as the response to a valid authorisation token request, which is an authorisation token generated by the authorisation server. The AISP requires the token to request PSU account data.

\begin{lstlisting}[language=AnBx]
    AuthToken: fAuthCodeToken(AuthorisationCode);
\end{lstlisting}

\noindent To generate the authorisation token, the authorisation server uses the function ~\verb'fAuthCodeToken'. The function accepts an authorisation code sent by the AISP. The code is validated and used to bind the token to the consent to which the code is mapped. As authorisation tokens are mapped to the corresponding consents, the resource server can ensure the AISP only accesses the permitted PSU data on the selected accounts.

\begin{lstlisting}[language=AnBx]
    Function [JWT -> JWT] fAuthCodeToken;
\end{lstlisting}

\noindent Note that there is no direct connection between AISPs and generated authorisation tokens. However, similar to the client tokens, the resource server is required to verify that an authorisation token used by an AISP belongs to the said AISP. As previously explained, we have a mapping between authorisation tokens and consents, and referring to substep 14 of the Account API Specification sequence diagram \cite{website:OB_security_profile_impl_draft}, we can see that consents are mapped to their AISPs. Consequently, to check ownership of an authorisation token, the resource server can check the consent linked to the token and then the AISP who created the consent. Similarly, to check ownership of an authorisation code, the mappings from codes to consents and consents to AISPs are used.

\parahead{Resource Server} \textit{Action A2.4} --~\verb'Intent' represents the response to a valid consent creation request, which is a consent resource created by the resource server.

\begin{lstlisting}[language=AnBx]
    Intent: fCreateIntent(ClientToken,IntentAgreement);
\end{lstlisting}

\noindent To create the consent resource, the resource server uses the \verb'fCreateIntent' function. As the first parameter, the function takes in the client token sent by and mapped to the AISP, to validate the token and map the created consent to the AISP. This mapping is based on sub-step 14 of the Account API Specification sequence diagram \cite{website:OB_security_profile_impl_draft}. The function also requires details of the consent to create as the second parameter. This is the consent agreement between the PSU and AISP, forwarded by the AISP to the resource server to create the consent.

\begin{lstlisting}[language=AnBx]
    Function [JWT, IntentNo -> JWT] fCreateIntent;
\end{lstlisting}

\noindent\textit{Action 3.2.2} -- ~\verb'RetrieveIntentRes' represents the response to a request to retrieve the consent resource to authorise and a list of PSU accounts. It consists of the requested consent resource and list of PSU accounts.

\begin{lstlisting}[language=AnBx]
    RetrieveIntentRes:fGetIntent(Intent),fFetchAccounts(PSU);
\end{lstlisting}

\noindent To retrieve these resources, the resource server is assumed to rely on two functions:~\verb'fGetIntent' and~\verb'fFetchAccounts'. The ~\verb'fGetIntent' function accepts a consent identifier and returns the corresponding consent resource. The ~\verb'fFetchAccounts' function accepts the identifier of the PSU whose accounts are to be returned. These identifiers are received as part of the request.

\begin{lstlisting}[language=AnBx]
    Function [JWT -> IntentNo] fGetIntent;
    Function [Agent -> AccountNumber] fFetchAccounts;
\end{lstlisting}

\noindent\textit{Action A3.2.6} -~\verb'AuthoriseIntentRes' serves as the response to a consent resource-authorisation request. It contains the success of updating the consent resource.

\begin{lstlisting}[language=AnBx]
    AuthoriseIntentRes: fAuthoriseIntent(PSU,Intent,SelectedAccounts);
\end{lstlisting}

\noindent To update a consent resource to an authorised state, the resource server uses a function named ~\verb'fAuthoriseIntent' in our model.

\begin{lstlisting}[language=AnBx]
    Function [Agent,JWT,IntentNo -> JWT] fAuthoriseIntent;
\end{lstlisting}

\noindent The function accepts three inputs. The first is the identifier of the PSU authorising the consent. This is used to link PSUs to the consents they authorise. This is not explicitly stated as a requirement in the specification, but it helps determine who controls the consents (e.g. revoke access). 

We assume that the consent is only linked to a PSU when it is authorised by the PSU and not when the consent resource is created. The reasoning is that when the AISP requests the creation of a consent, they are not required to send information identifying the PSU with whom the consent is associated. However, during the authorisation process, the PSU is required to send their credentials to the authorisation server to be authenticated. The authorisation server can then forward the PSU's identifier from the credentials to the resource server.

The second input to the function is the identifier of the consent to authorise. The third input is the set of PSU accounts that the PSU has selected to associate with the consent. This should be a subset of the accounts obtained by the resource server through the~\verb'fFetchAccounts' function and sent to the PSU to choose from. This input is required as the~\verb'fAuthoriseIntent' function is assumed to also link the selected accounts to the corresponding consent. This allows the resource server to determine which accounts to return data to the AISP. The function output determines the success of authorising the consent.

The data for each of the parameters are received as part of the request to authorise the consent resource.

\vspace{5pt}

\noindent\textit{Action A4.2} --~\verb'AccountsRes' represents the response to a valid PSU account data retrieval request. It includes ~\verb'Accounts', which is the requested permitted general PSU account data. As previously discussed, we do not model here the scenario where a PSU does not select any account.

\begin{lstlisting}[language=AnBx]
    Accounts: fAccountsEndpoint(AuthToken,AccountsSecretNonce,empty);
\end{lstlisting}

\noindent To retrieve the general PSU account data, the resource server relies on the ~\verb'fAccountEndpoint' function, which takes three inputs. It accepts an authorisation token sent by the AISP. The token is validated to ensure that the AISP is allowed to access the data. It is assumed that the token is also used to obtain the corresponding consent, so the resource server only provides the AISP with the data on the accounts the PSU has agreed on. The second parameter is the unique id (UUID) used to distinguish between different requests. The third parameter determines whether to return data on all allowed accounts (bulk call) or on a specific account (specific call). If it is \texttt{\small\textbf{empty}}, it is a bulk call; otherwise, it is a specific call.

\begin{lstlisting}[language=AnBx]
    Function [JWT,UUID,Number -> AccountNumber] fAccountsEndpoint;
\end{lstlisting}

\noindent As illustrated in Section~\ref{sec:background}, there are many endpoints to request different PSU data. However, modelling all possible endpoints is unnecessary in \textit{AnBx} as the required authentication/authorisation procedure has already been completed. This is because the main distinction between those endpoints is the data models of the data they return, which is unsuitable to be modelled using \textit{AnBx} and can be abstracted. Therefore, we only model one of the aforementioned endpoints.

In Section \ref{sub:atp}, specifically Step 4 of the Account and Transaction API flow, a requirement of the specification is that the AISP must request the list of PSU accounts as the first request with a new authorisation token. Based on this requirement, it was decided to model the endpoint for accessing general PSU account data, which also returns the list of permitted accounts. This functionality is represented by the~\verb'fAccountsEndpoint' function previously explained.

\section{Automated Verification}
\label{sec:security}

In this section, we describe the security goals considered in the modelling and  verification processes.

\subsection{Security Goals}\label{sub:goals}

The primary objective of our formal model is to verify whether the protocol satisfies specific security properties (goals), using automated verification tools. These goals are derived from our interpretation of the protocol, its dependencies, and related security considerations such as OAuth 2.0 security requirements \cite[P.52-P.60]{report:rfc6749_auth-2.0}, and use cases outlined in the Open Banking standard \cite[P.20-P.23]{report:OB_standard}, despite no explicit goals are stated within the ATP requirements \cite{report:OB_standard}.

We have identified eight security goals, categorised into four goals on message confidentiality and four on authentication, which we believe capture essential security properties expected for this protocol.

\begin{lstlisting}[language=AnBx]
 fAISPSecret(aspspA,AISP) secret between AISP,aspspA         #G1
 fPSUSecret(PSU,aspspA)   secret between PSU,aspspA          #G2
 ClientToken,AuthToken    secret between AISP,aspspA,aspspR  #G3

 PSU    authenticates aspspA on fGetIntent(Intent)           #G4
 aspspR weakly authenticates PSU on SelectedAccounts         #G5
 PSU authenticates AISP on ASPSPAuthPSUEndpoint,AISPEndpoint #G6

 Accounts secret between AISP,aspspR                         #G7
 AISP     weakly authenticates aspspR on Accounts            #G8
\end{lstlisting}

Goals \emph{G1} and \emph{G2} ensure that secrets and credentials exchanged between AISP, PSU, and authorisation server (aspspA) remain confidential during actions, such as requesting a client token (\emph{A2.1}) and obtaining consent authorization (\emph{A3.1.2} and \emph{A3.3.3}). This confidentiality is crucial, as the leakage of AISP or PSU credentials could lead to severe security risks, including client impersonation \cite[Sect.~10.2]{report:rfc6749_auth-2.0}.

Goal \emph{G3} requires that various tokens exchanged (\emph{A2.2--A2.3} and \emph{A3.3.4--A4.1}) between the AISP and the authorisation/resource servers (aspspA, aspspR) remain confidential. This prevents attacks involving token injection and aligns with OAuth 2.0's requirement for token confidentiality \cite[Sect.~10.3, 10.12]{report:rfc6749_auth-2.0}. These goals clearly indicate the inherent potential vulnerabilities of the ATP dependencies.

The final secrecy goal \emph{G7} ensures that the account information exchanged between the AISP and the resource server (aspspR) (\emph{A4.2}) remains secret, hence protecting sensitive financial data.

Authentication goals relate to the PSU authenticating the authorisation server on the consent resource to authorise (\emph{G4}), the resource server authenticating the PSU's selected account information (\emph{G5}), and the AISP authenticating the PSU's account information from the resource server (\emph{G8}). It should be noted that the kind of authentication expected in \emph{G5} and \emph{G8} is non-injective (i.e. weak authentication), as the accounts retrieved can be the same on different runs of the protocols.

Similarly, goal \emph{G6} ensures that the PSU successfully authenticates the AISP during interactions with specific endpoints (\texttt{\small{ASPSPAuthPSUEndpoint}} and \texttt{\small{AISPEndpoint}}). This authentication is critical to verify the identity of the AISP before proceeding with sensitive operations, maintaining trust and security throughout the protocol.

\subsection{Verification}

\begin{table}
	\centering
	\begin{tabular}{>{\centering\arraybackslash}p{0.08\textwidth}|>{\centering\arraybackslash}p{0.10\textwidth}|>{\centering\arraybackslash}p{0.14\textwidth}|>{\centering\arraybackslash}p{0.14\textwidth}|>{\centering\arraybackslash}p{0.14\textwidth}|>{\centering\arraybackslash}p{0.14\textwidth}}
		\textit{} & \textit{Goal} & \textit{OFMC} & \textit{OFMC} & \textit{OFMC} & \textit{ProVerif} \\
		\textit{Goal} & \textit{Type} & \textit{1 session} & \textit{2 sessions} $d=12$ & \textit{2 sessions} $d>12$ & \textit{unlimited sessions} \\
		\hline
		\textit{G1} & $Sec$ & \checkmark & \checkmark & \textit{timeout} & \checkmark \\
		\textit{G2} & $Sec$ & \checkmark & \checkmark  & \texttt{"} & \checkmark \\
		\textit{G3} & $Sec$ & \checkmark & \checkmark & \texttt{"} & \checkmark \\
		\textit{G4} & $Auth$& \checkmark & \checkmark & \texttt{"} & $WAuth$\\
		\textit{G4} & $WAuth$ & \checkmark & \checkmark & \texttt{"} & \checkmark\\
		\textit{G5} & $WAuth$ & \checkmark & \checkmark & \texttt{"} & \checkmark \\
		\textit{G6} & $WAuth$ & \checkmark & \checkmark & \texttt{"} & \checkmark \\
		\textit{G7} & $Sec$ & \checkmark & \checkmark & \texttt{"} & \checkmark \\
		\textit{G8} & $WAuth$ & \checkmark & \checkmark & \texttt{"} & \checkmark \\
		\hline
		\textit{time} & \textit{all} & $13s$ & $12~days$ & $>12~days$ & $3m25s$ \\      
        \textit{time} & \textit{single} & -- & --  & -- & $1m18s$ \\      
	\end{tabular}
	
	\vspace{2mm}
	\small $Sec$ = Secrecy, $Auth$ = Authentication, $WAuth$ =  Weak Authentication
	
	\caption{Verification Results -- \textit{AnB} (OFMC) and \textit{Applied-pi} (ProVerif) Typed Models}
	\label{tab:verification}
\end{table}

The verification was carried out on an Intel(R) Xeon(R) W-2145 CPU @ 3.70GHz with 128GB of RAM running Windows 10 64-bit. We began by using the OFMC model checker \cite{basin2005ofmc} (version 2022) to verify the eight security goals described in Section \ref{sub:goals}, including injective authentication (\emph{G4}).  All goals were successfully verified for a single session, with the overall verification completing in approximately 13 seconds.

However, due to the potential for attacks exploiting interleaved sessions between agents, verifying the model for only one session is insufficient. The state search space increases drastically with multiple sessions, which poses an insurmountable challenge to exhaustive verification within our hardware constraints.

As is customary in such situations (e.g., \cite{DBLP:journals/istr/BugliesiCMM16} for iKP and SET), we achieved partial results by setting the number of sessions to 2 and expanding the search space up to the available resource limits (namely, a depth $d$ of 12 plies, running for 286 hours, almost 12 days) without reaching any attack state. Increasing $d$ by just one ply would have resulted in an estimated 7–9 times longer verification time. This estimation is based on previous partial results at lower depths, specifically our observations when increasing $d$ from 11 to 12.  

With 4 agents and 19 protocol steps, the expected maximum number of plies could be at least 76. This suggests that we were far from fully verifying the model for two sessions with OFMC. In other words, even with greater computational power, given the exponential growth of each additional ply, OFMC verification becomes infeasible.

To enable investigating verification of the model for multiple sessions, we used ProVerif version 2.05 \cite{Blanchet2022}. The \textit{AnBx} compiler is capable of generating an applied pi model from the same \textit{AnBx} model that we used for OFMC. However, we needed to expand the bullet channels with a concrete cryptographic implementation that conveys the same security goals in order to generate a model verifiable by ProVerif, as the tool does not support such channels directly. The soundness of the channel translation in a concrete cryptographic implementation has been proved in \cite{DBLP:journals/istr/BugliesiCMM16}, being the bullet channels a specific instance of the more general \textit{AnBx} channels. It should also be noted that although in ProVerif there is a notion of private channel, this is not equivalent to the OFMC secure channel, as in ProVerif the intruder cannot interact with or observe private channel communication. %

Specifically, we used an implementation of bullet channels where using asymmetric encryption for secrecy and digital signature for authenticity. In this case, each step of a protocol generates a single step with the explicit application of cryptography for a efficient translation strategy.

For example, a secure channel is represented by the bullet channel notation as \verb'A*->*B: Msg' in \textit{AnBx}. This is syntactic sugar that is mapped to \verb'A->B,(A|B|B): Msg' in the \textit{AnBx} channel notation \cite{DBLP:journals/istr/BugliesiCMM16}, and finally to:

\begin{lstlisting}[language=AnBx]
	A->B: {{B,Msg}inv(sk(A))}pk(B)
\end{lstlisting}

\noindent where \verb'B' is the identity of the intended recipient, \verb'Msg' is the fresh message, and the payload is first digitally signed by the sender \verb'A' with its private key \verb'inv(sk(A))' and then encrypted with \verb'pk(B)', the public key of \verb'B'.

In essence, we model an abstract channel capable of achieving security goals, since such assumption can be translated to a concrete cryptographic implementation. The compilation of \textit{AnBx} channels (a generalisation of \textit{AnB} bullet channels) and the soundness of the translation are detailed in \cite{DBLP:journals/istr/BugliesiCMM16}.

With ProVerif, we were able to verify all goals for an unlimited number of sessions. The only limitation was that for goal \emph{G4}, we were only able to prove weak authentication, since for strong authentication, the verification process ran out of memory after 6 hours. 

Results are summarised in Table~\ref{tab:verification}. The verification with two sessions proved to be beyond the capability of OFMC, but ProVerif was able to verify the model for an unbounded number of sessions in less than 3 minutes and 30 seconds. Interestingly, the verification time was further reduced to just 1 minute and 18 seconds when testing the security goals individually in parallel. This is a feature enabled by the AnBx IDE \cite{anbx-ide2024}, which orchestrates the verification of single-goal versions of the protocol under test, that are generated by the AnBx compiler from the original model.

\section{Testing an Open Banking Sandbox}
\label{sec:sandbox-testing}

As discussed by Kellezi et al. \cite{kellezi2021securing}, developers must register with and be approved by national financial authorities to use Open Banking APIs in a production environment. This poses a significant barrier to independent investigation of concrete Open Banking system implementations. Therefore, using a sandbox is the only practical option for independent researchers.

NatWest, one of the largest banks in the UK, released the first production version of their Open Banking Account and Transaction API v3.1.11 in April 2024 \cite{natwest}. Together with a few other banks, they also released a sandbox version, allowing potential third-party providers (TPPs) to use the API in a test environment.

It should be noted that the NatWest sandbox environment, like any other sandbox, has limitations compared to its production environment. Although it provides essential functionalities, it does not fully replicate the scalability, performance, or certain security measures of the actual production environment. Additionally, since the sandbox essentially allows developer access, not all protocol steps can be audited during the process of testing the API. Some steps involve backend operations and network interactions not directly visible in the user interface, such as interactions with the ASPSP Authorisation and Resource servers.

However, having access to the NatWest sandbox allowed us not only to gain a better understanding of the system for the purpose of building the formal model, but also to examine some core security mechanisms of the implementation.

We started by running the Account and Transaction API protocol in the sandbox environment, including login requests and responses. In addition to collecting data for subsequent tests, the information was used to validate the previously developed formal model, ensuring that it is a realistic, albeit abstract, representation of the system.

We designed relevant sanity checks as security test cases implemented within the sandbox (see Section~\ref{sub:security-tests} and Figure~\ref{fig:anb_aisp_seq_diagram}) terms and conditions. These (19) test cases, tagged \textit{opnbnk01} through \textit{opnbnk19}, were spread over eight different test areas. The tests were carried out manually using Postman \cite{postman}. Furthermore, we checked the properties of the access token (e.g. integrity, freshness and scope), as this is a crucial mechanism in the implementation of the protocol to ensure the integrity and authenticity of the token payload (Section~\ref{sub:access-token}).

\subsection{Security Tests}
\label{sub:security-tests}

As reported in Table~\ref{tab:testscases}, manual testing was performed to interact with the endpoints and observe the API's behaviour and functionality. The protocol steps correspond to those depicted in Figure~\ref{fig:anb_aisp_seq_diagram}, with all $19$ tests yielding expected results and confirming successful execution. That is, each test correspond to parts of the diagram which are executable within the sandbox.
  
These tests empirically demonstrate the system's capability in maintaining account data integrity and enforcing access restrictions, specifically in detecting and preventing unauthorised access attempts. However, it is important to acknowledge the limitations of this approach, as these tests do not constitute a comprehensive investigation of the security properties of the NatWest implementation. 
Manual testing is inherently constrained by its scope and scale. Additionally, these tests are conducted within the legal boundaries set by current legislation and the sandbox's terms and conditions. For instance, the prohibited uses include generating excessive calls, reverse engineering, and executing any malicious activities.

Therefore, such tests serve as essential security sanity checks, complementing the formal model verification presented in previous sections. In fact, in an actual implementation the various assumptions and preconditions of each protocol step must be satisfied, i.e. they are necessary conditions. This implies that, if the tests fail, then some of the \textit{AnBx} security goals will fail as well. On the other hand, the tests alone are not sufficient to establish the security goals from the formal model within the sandbox implementation.

We first describe the scenarios and key objectives of the $19$ security tests, listed in Table~\ref{tab:testscases}, and later describe them in detail. Test \textit{opnbnk01} focuses on data validation, specifically examining network layer security, to prevent clients from interacting with the server via unsecured connections, thereby reducing the risk of data interception or tampering.

Tests \textit{opnbnk02} and \textit{opnbnk03} involve retrieving authorisation tokens, verifying whether PSUs could generate tokens using valid client secret keys.

In \textit{opnbnk04} and \textit{opnbnk05}, account consent requests are scrutinised to ensure that users cannot initiate requests with invalid authorisation tokens, and unauthorised users cannot initiate requests at all.

Tests \textit{opnbnk06}, \textit{opnbnk07}, and \textit{opnbnk08} ensure that account consent requests could not proceed with invalid client IDs, PSU usernames, or consent IDs, thereby effectively preventing unauthorised access attempts.

Tests \textit{opnbnk09}, \textit{opnbnk10}, and \textit{opnbnk11} focus on validating the request for new authorisation tokens. Checks were conducted to prevent the generation of new tokens using invalid authorisation codes, expired codes, or incorrect PSU usernames and client secrets/IDs. These measures enhance the system's security by thwarting unauthorised access attempts and maintaining the integrity of the authorisation process.

Data retrieval operations, such as fetching account lists and details, are validated by tests \textit{opnbnk12} through to \textit{opnbnk19}. These tests include checks to ensure that invalid API access tokens or account identifiers do not grant unauthorised access to sensitive information from other banks, hence safeguarding user privacy and preventing unauthorised data exposure.

The security tests primarily focus on negative testing, ensuring that the system correctly rejects invalid inputs and prevents unauthorised access. These tests verify the system’s resilience against security threats by simulating failure scenarios, such as incorrect authorisation tokens, expired credentials, and unauthorised data retrieval attempts. However, certain tests also include positive testing aspects, implicitly confirming that valid inputs produce the expected successful outcomes. For instance, in tests \textit{opnbnk02} and \textit{opnbnk03}, while the primary focus is on rejecting invalid credentials, the process also verifies that a valid client ID and secret key successfully generate an authorisation token.

\begin{sidewaystable}
	\caption{Security Test Cases}\label{tab:testscases} 
	\centering
	\small
\makebox[\textwidth][c]{
	\begin{tabular}{|>{\centering\arraybackslash}p{12mm}|>{\centering\arraybackslash}p{12mm}|>{\raggedright\arraybackslash}p{24mm}|>{\raggedright\arraybackslash}p{32mm}|>{\raggedright\arraybackslash}p{24mm}|>{\raggedright\arraybackslash}p{40mm}|>{\raggedright\arraybackslash}p{32mm}|}
		\hline 
		\textit{Test Case ID}  & \textit{Prot. Step}  & \textit{Test Case Scenario} & \textit{Test Case}  & \textit{Pre-Conditions} & \textit{Test Steps on Postman} & \textit{Test Results} \\
		\hline 
		\textit{opnbnk} 01  & Before \textit{A1.1} & Network Layer Validations  & Client cannot interact with the server using unsecured network connection.  & Network connection must be http  & \smallurl{https://api.sandbox.natwest.com/.well-known/openid-configuration} Enter invalid authorisation token & Request was not honoured as expected \\
		\hline 
		\textit{opnbnk} 02 03  & \textit{A2.1} & Retrieve authorisation token  & PSU cannot generate authorisation token using invalid client ID and
		secret key  & client ID or secret key must be invalid  & \smallurl{https://ob.sandbox.natwest.com/token} \newline
		Enter invalid client ID and secret   & Authorisation token request was unsuccessful as expected \tabularnewline
		\hline 
		\textit{opnbnk} 04 05  & \textit{A3.1} & Account consent request  & PSU or unauthorised user cannot initiate an account consent request
		using invalid authorisation token  & Authorisation token must be invalid  & \smallurl{https://ob.sandbox.natwest.com/open-banking/v3.1/aisp/account-access-consent}
		\newline
		Enter invalid authorisation token & Account consent request was declined as expected \tabularnewline
		\hline 
		\textit{opnbnk} 06 07 08  & \textit{A3.1.2} & Approve account consent request  & Account consent request cannot be approved using invalid client ID,
		invalid PSU username or consent id  & client ID, PSU username and consent ID must be invalid  & \smallurl{https://ob.sandbox.natwest.com/open-banking/v3.1/aisp/account-access-consents}
		\newline
		Enter invalid authorisation token & Consent request approval was declined as expected \tabularnewline
		\hline 
		\textit{opnbnk} 09 10 11  & \textit{A3.2.3} & Request new authorisation token  & New authorisation token cannot be generated using invalid, empty,
		or expired authorisation code  & Authorisation code must be invalid, empty or expired  & \smallurl{authorisationEndpoint} \newline
		Enter invalid authorisation token   & New authorisation token request was unsuccessful as expected \tabularnewline
		\hline 
		\textit{opnbnk} 12 13  & \textit{A4.1} & Request list of account using invalid data  & List of accounts cannot be fetched using invalid API access token  & Access token must be invalid  & \smallurl{https://ob.sandbox.natwest.com/token} \newline
		Enter invalid authorisation token   & Request for list of accounts was declined as expected \tabularnewline
		\hline 
		\textit{opnbnk} 14 15 16  & \textit{A4.1} & Get single account details using invalid data  & Account details cannot be fetched using invalid account number, invalid
		API access token  & Account number or access token must be invalid  & \smallurl{{apiUrlPrefix}/open-banking/v3.1/aisp/accounts} \newline
		Enter invalid authorisation token   & Fetching single account details Using invalid data failed as expected \tabularnewline
		\hline 
		\textit{opnbnk} 17 18 19  & \textit{A4.1} & Get transactions using invalid data  & Get transactions using invalid account ID, unauthorised or API access
		token  & API access token or acct no must be invalid  & \smallurl{apiUrlPrefix} \newline
		Enter invalid authorisation token   & Request for transactions failed as expected \tabularnewline
		\hline

	\end{tabular}
}
\end{sidewaystable}

\subsubsection{Network Layer Validations}
Test \textit{opnbnk01} assess if the PSUs and AISPs could communicate with the server via HTTP to access a secure endpoint, which also authenticates the PSU before Action \textit{A1.1}. The system correctly identifies the unsecure connection and reject requests, ensuring compliance with security guidelines and mitigating risks from unsecure connections.

\subsubsection{Retrieve Authorisation Token}
Tests \textit{opnbnk02} and \textit{opnbnk03} verify the security of authorisation token retrieval (\textit{A2.1}). Test \textit{opnbnk02} check that the tokens could not be generated with an invalid client ID, while \textit{opnbnk03} ensure tokens could not be generated with an invalid client secret key. Using Postman, invalid credentials are sent to the token retrieval URL, and both tests confirm the system's ability to detect and prevent token forging.

\subsubsection{Account Consent Request}
Tests \textit{opnbnk04} and \textit{opnbnk05} ensure the security of initiating account consent requests. Test \textit{opnbnk04} verify that AISPs cannot initiate requests with an invalid authorisation token (\textit{A3.1}), while test \textit{opnbnk05} prevents unauthorised users from initiating requests, even with invalid tokens (\textit{A3.1}). Using Postman, invalid tokens are sent and the system correctly declined requests, demonstrating its ability to prevent unauthorised access.

\subsubsection{Approve Account Consent Request}
Tests \textit{opnbnk06}, \textit{opnbnk07}, and \textit{opnbnk08} (\textit{A3.1.2}) validate the approval process for account consent requests. Test \textit{opnbnk06} confirms that consent requests cannot be approved using invalid client IDs, PSU usernames, consent IDs, or authorisation tokens. Tests \textit{opnbnk07} and \textit{opnbnk08} show the rejection of approval attempts with invalid PSU usernames and consent IDs, respectively. The expected results demonstrated effective validation checks for individual tests \textit{opnbnk06}, \textit{opnbnk07}, and \textit{opnbnk08}.

\subsubsection{Request New Authorisation Token}
Tests \textit{opnbnk09}, \textit{opnbnk10}, and \textit{opnbnk11} examine system's vulnerabilities to generating new authorisation tokens with invalid, empty, or expired authorisation codes. Using Postman, test \textit{opnbnk09} sends a request with an invalid authorisation code, resulting in an \texttt{invalid\_grant} error (\textit{A3.2.3}). Test \textit{opnbnk10} submits an empty authorisation code parameter, receiving an \texttt{invalid\_request} error. Test \textit{opnbnk11} use an expired authorisation code, with the server correctly responding with an \texttt{invalid\_grant} error.

\subsubsection{Request List of Accounts Using Invalid Data}
Tests \textit{opnbnk12} and \textit{opnbnk13} evaluate the system's response to account list requests with invalid or unauthorised access tokens (\textit{A4.1}). Test \textit{opnbnk12} confirms the system's rejection of unauthorised access with an invalid token. Test \textit{opnbnk13} assess the system's exclusion of accounts from banks other than the designated one using an invalid permission token, producing the expected negative response and demonstrating adherence to access parameters (\textit{A4.2}).

\subsubsection{Get Single Account Details Using Invalid Data}
Tests \textit{opnbnk14}, \textit{opnbnk15}, and \textit{opnbnk16} evaluate the system's handling of single account information retrieval with unauthorised or incorrect data (\textit{A4.1}). Test \textit{opnbnk14} confirms the system's correct rejection of a fictitious account number. Test \textit{opnbnk15} demonstrates the system's denial of requests with invalid API access tokens. Test \textit{opnbnk16} validate the exclusion of account details not affiliated with the specified bank, ensuring data integrity and access parameters adherence.

\subsubsection{Get Transactions Using Invalid Data}
Tests \textit{opnbnk17}, \textit{opnbnk18}, and \textit{opnbnk19} assess transaction retrieval requests involving invalid or unauthorised data (\textit{A4.1}). Test \textit{opnbnk17} confirms the system's enforcement of data integrity by rejecting invalid account IDs. Test \textit{opnbnk18} demonstrates the detection and rejection of unauthorised access attempts with invalid API tokens. Finally, test \textit{opnbnk19} shows the system's adherence to access parameters by systematically denying unauthorised user access to transaction data.

\subsection{Access Tokens}
\label{sub:access-token}

Throughout the protocol execution, various tokens were obtained via HTTP POST requests to the NatWest Open Banking sandbox. Figure \ref{fig:DecodedAccessToken} (Action \textit{A2.2}) depicts the decoded response from the OpenID Connect configuration endpoint, which serves as the initial point of contact. Figure \ref{fig:DecodedAccessTokensA3.3.4} (Action \textit{A3.3.4}) shows the decoded response from the authorisation endpoint after redirecting a PSU for authorisation.

\begin{figure}[t]
  \centering
  \begin{lstlisting}[language=json]
  HEADER:ALGORITHM & TOKEN TYPE
{
  "alg": "RS256",
  "typ": "JWT"
}
  PAYLOAD:DATA
{
  "app": "Qudus App",
  "org": "linkfieldcare.com",
  "iss": "https://api.sandbox.natwest.com",
  "token_type": "ACCESS_TOKEN",
  "external_client_id": "eJsuhFdtK6aJEtpVO08cywoAXvX1icCQJmVz3Qou3kk=",
  "client_id": "7170f9f0-c496-4b4b-b6e7-50c205fe1c09",
  "max_age": 86400,
  "aud": "7170f9f0-c496-4b4b-b6e7-50c205fe1c09",
  "scope": "accounts",
  "exp": 1713863496,
  "iat": 1713862896,
  "jti": "e79d9032-a92e-415e-8845-73787e0abc21",
  "tenant": "NatWest"
}
\end{lstlisting}
  \caption{Decoded Access Token of Action \textit{A2.2} 
  	} \label{fig:DecodedAccessToken}
\end{figure}

\begin{figure}[th!]
  \centering

  \begin{lstlisting}[language=json]
   ACCESS TOKEN - HEADER:ALGORITHM & TOKEN TYPE
{
  "alg": "RS256",
  "typ": "JWT"
}
  ACCESS TOKEN - PAYLOAD:DATA
{
  "app": "Qudus App",
  "org": "linkfieldcare.com",
  "iss": "https://api.sandbox.natwest.com",
  "token_type": "ACCESS_TOKEN",
  "external_client_id": "eJsuhFdtK6aJEtpVO08cywoAXvX1icCQJmVz3Qou3kk=",
  "client_id": "7170f9f0-c496-4b4b-b6e7-50c205fe1c09",
  "max_age": 86400,
  "aud": "7170f9f0-c496-4b4b-b6e7-50c205fe1c09",
  "user_id": "123456789012@linkfieldcare.com",
  "grant_id": "77cdac57-5321-422a-b7a0-3d3affcd0515",
  "scope": "accounts openid",
  "consent_reference": "8d0ab30c-2b99-48f3-81dc-d1a1aef63b3e",
  "exp": 1714479564,
  "iat": 1714478964,
  "jti": "ec2e6050-8588-4ee2-9c5e-adfd947113d7",
  "tenant": "NatWest"
}
   REFRESH TOKEN - HEADER:ALGORITHM & TOKEN TYPE
{
  "alg": "RS256",
  "typ": "JWT"
}
  REFRESH TOKEN - PAYLOAD:DATA
{
  "app": "Qudus App",
  "org": "linkfieldcare.com",
  "iss": "https://api.sandbox.natwest.com",
  "token_type": "REFRESH_TOKEN",
  "external_client_id": "eJsuhFdtK6aJEtpVO08cywoAXvX1icCQJmVz3Qou3kk=",
  "client_id": "7170f9f0-c496-4b4b-b6e7-50c205fe1c09",
  "max_age": 86400,
  "aud": "7170f9f0-c496-4b4b-b6e7-50c205fe1c09",
  "user_id": "123456789012@linkfieldcare.com",
  "grant_id": "77cdac57-5321-422a-b7a0-3d3affcd0515",
  "scope": "accounts openid",
  "consent_reference": "8d0ab30c-2b99-48f3-81dc-d1a1aef63b3e",
  "iat": 1714478964,
  "jti": "90afc92a-567c-4795-a4c2-731308a2b2a6",
  "tenant": "NatWest"
}
  ID-TOKEN - HEADER:ALGORITHM & TOKEN TYPE
{
  "alg": "PS256",
  "typ": "JWT",
  "kid": "wFZlD45hv45gOsq8Dac3lxw5FpA"
}
  ID-TOKEN - PAYLOAD:DATA
{
  "sub": "c91b0fc1-f9ab-4e54-8403-7298f5cb60e6",
  "acr": "urn:openbanking:psd2:sca",
  "aud": "eJsuhFdtK6aJEtpVO08cywoAXvX1icCQJmVz3Qou3kk=",
  "c_hash": "ohWEFcvxK-8uaaSVzMcaMQ",
  "openbanking_intent_id": "c91b0fc1-f9ab-4e54-8403-7298f5cb60e6",
  "s_hash": "tdQEXD9Gb6kf4sxqvnkjKg",
  "auth_time": 1714478959,
  "iss": "https://api.sandbox.natwest.com",
  "exp": 1746014959,
  "token_type": "ID_TOKEN",
  "iat": 1714478959,
  "jti": "c5792bdc-7a32-4514-b852-088dbed8e1ce"
}
\end{lstlisting}
  \caption{Decoded Authorisation Tokens of Action \textit{A3.3.4}
  }\label{fig:DecodedAccessTokensA3.3.4}
\end{figure}

A JSON Web Token (JWT) is a widely used representation format for token exchange. It consists of three parts: a header, a payload, and a signature, computed as follows:
\[
X = \text{Base64URLEncode}(\textit{header}) + '.' + \text{Base64URLEncode}(\textit{payload})
\]
\[
\text{JWT} = X + '.' + \text{Base64URLEncode}(\textit{sigalg}(X))
\]
\noindent   where \textit{sigalg} is the chosen signing algorithm, and Base64URLEncode is URL-safe format using base 64 encoding. %
These tokens are encoded and typically signed with the RS256 algorithm (RSA with SHA-256), a widely adopted standard for JWT signatures, ensuring the integrity and authenticity of the token payload. The token should be validated by the recipient, as specified in Section 7.2 of RFC 7519 \cite{rfc7519}, and then decoded. Next, the recipient verifies the digital signature and, if successful, processes the payload. Figures \ref{fig:DecodedAccessToken} and \ref{fig:DecodedAccessTokensA3.3.4} show sample decoded tokens, omitting the signature for brevity. Note that not all fields described are always present in all tokens.

The \textit{header} specifies the cryptographic algorithm and token type:

\begin{longtable}{lp{10.4cm}}
	\textit{Field} & \textit{Description} \\ \hline
	\endfirsthead
	\multicolumn{2}{c}%
	{\tablename\ \thetable\ -- \textit{Continued from previous page}} \\
	\textit{Field} & \textit{Description} \\ \hline
	\endhead
	\multicolumn{2}{r}{\textit{Continued on next page}} \\
	\endfoot
	\endlastfoot
	\textit{alg} & RS256 (RSA with SHA-256) for signing the token, ensuring integrity and authenticity \\ %
	\textit{typ} & JWT (JSON Web Token) for secure information transmission \\ %
	\textit{kid} & Optional key ID for signalling key changes to recipients \cite{rfc7515} \\ 
\end{longtable}

\noindent The \textit{payload} contains claims and context-related information about the token's origin, purpose, audience, permissions, and validity:

\begin{longtable}{lp{8.3cm}}
	\textit{Field} & \textit{Description} \\ \hline
	\endfirsthead
	\multicolumn{2}{c}%
	{\tablename\ \thetable\ -- \textit{Continued from previous page}} \\
	\textit{Field} & \textit{Description} \\ \hline
	\endhead
	\multicolumn{2}{r}{\textit{Continued on next page}} \\
	\endfoot
	\endlastfoot
	\textit{app} & Application name associated with the token \\ %
	\textit{org} & Organisation associated with the token \\
	\textit{iss} & Issuer of the token, typically the identity provider's URL (NatWest API sandbox) \\ %
	\textit{token\_type} & Type of token, such as an access token \\ %
	\textit{external\_client\_id} & External client ID linked to the token \\ %
	\textit{max\_age} & Maximum allowable token age in seconds (86,400 or 24 hours)
	 \\ %
	\textit{aud} & Audience intended to receive the token \\ %
	\textit{scope} & Scope of access granted by the token \\ %
	\textit{exp} & Expiration time as a Unix timestamp \\ %
	\textit{iat} & Issuance time as a Unix timestamp \\ %
	\textit{jti} & Unique JWT ID \\ %
	\textit{client\_id} & Audience or client ID of the token \\ %
	\textit{user\_id} & Unique user identifier associated with the token \\ %
	\textit{grant\_id} & Specific authorisation grant tied to the token \\
	\textit{consent\_reference} & Unique reference for user consent \\ %
	\textit{tenant} & Tenant associated with the token, representing a specific entity within ASPSP \\ %
	\textit{sub} & Subject, the unique user identifier \\ %
	\textit{acr} & Authentication context class for user authentication, compliant with PSD2 Strong Customer Authentication (SCA) as \texttt{urn:openbanking:psd2:sca} \\ %
	\textit{hash values} & Cryptographic hashes (\textit{c\_hash} and \textit{s\_hash}) for token validation \\ %
	\textit{intent\_id} & Unique identifier (\textit{openbanking\_intent\_id}) for a specific intent tied to the token \\ %
	\textit{auth\_time} & User authentication time (Unix timestamp) \\ 
\end{longtable}

The security measures within each token enhance the robustness of the authentication and authorisation processes. %
Specifically, the JWT ID (\texttt{jti}) uniquely identifies each token, preventing replay attacks and aiding audits. Tokens include an expiration time (\texttt{exp}) and an issued-at timestamp (\texttt{iat}), enforcing temporal constraints to mitigate unauthorised access. The issuer (\texttt{iss}) and audience (\texttt{aud}) claims ensure the token's authenticity and its intended use, preventing misuse. Token properties evolve during refresh cycles, maintaining freshness and ensuring access is restricted to defined timeframes.

The examination of provided tokens shows differences in scope and intended use. The Access Token in Figure \ref{fig:DecodedAccessToken} is a JWT signed with RS256, scoped to accounts, and lacks user-specific identifiers like \texttt{user\_id} and \texttt{grant\_id}. This token is suitable for account management operations without requiring OpenID Connect functionalities.

In contrast, the Authorisation Token in Figure \ref{fig:DecodedAccessTokensA3.3.4} has a broader scope and includes \texttt{user\_id}, \texttt{grant\_id}, and \texttt{consent\_reference}. This token is appropriate for accessing both account data and OpenID Connect services.
The Refresh Authorisation Token in Figure \ref{fig:DecodedAccessTokensA3.3.4} (Action \textit{A3.3.4}) allows obtaining a new Access Token without reauthentication, extending session validity and ensuring PSU legitimacy.

In details, the Authorisation Token comprises three essential sub-components: an access token, a refresh token, and an ID token.

\begin{itemize}[noitemsep]
	\item \textit{Access Token}: Grants temporary access to protected resources on behalf of the PSU. It contains user permissions and scopes that allow back-end services to make access control decisions. Short-lived and transmitted securely over HTTPS, it uses the RS256 algorithm for digital signatures. Key attributes include issuer (\texttt{iss}), token type (\texttt{token\_type}), expiration time (\texttt{expires\_in}), and access scope (\texttt{scope}).
	
	\item \textit{Refresh Token}: Acquires a new access token without PSU reauthentication once the current token expires. Longer-lived and securely stored, refresh tokens are bound to specific clients and scopes, ensuring continuous access and enhancing security and usability.
	
	\item \textit{ID Token}: Used in OpenID Connect flows for user authentication and authorisation. Contains user information such as user ID, authentication time, and issuer details. Digitally signed for integrity, ID tokens include a nonce to prevent replay attacks and facilitate single-sign-on (SSO) across applications. In this particular instance the PS256 algorithm (probabilistic RSA Signature with SHA-256) is used to ensure authenticity and integrity.
\end{itemize}

As we will discuss in Section \ref{sec:relatedwork}, the properties of such tokens and their security implications have been extensively studied by Fett et al.~\cite{fett2019extensive}, and other researchers (e.g., \cite{Mainka2017, Navas2019,Philippaerts2022, Innocenti2023}). Therefore, further investigation of such aspects is beyond the scope of our work. %

\section{Discussion and Limitations}
\label{sec:evaluation}
In general, formal modelling and verification offer a comprehensive and mathematically rigorous analysis of protocol security. This approach requires specifying a model of the protocol and its security goals abstractly enough to enable  employing automated tools to verify these goals across all possible scenarios defined by the attacker's capabilities \cite{Kulik2022}. Tools such as OFMC and ProVerif enable discovery about the presence of attacks in the Dolev-Yao intruder model \cite{dolev83ieee}. It is important to note that while OFMC is both sound and complete, ProVerif is sound but not complete.
However, in our case, since ProVerif did not detect attacks, false positives are not a concern. In fact, due to ProVerif soundness when a property is satisfied, then the model guarantees that property. The only limitation with ProVerif was that for goal \emph{G4}, we were only able to prove weak authentication, as with strong authentication, the verification process ran out of memory. In contrast, with OFMC, we were able to complete the verification process including strong authentication goals, but for one session only.

The effectiveness of formal methods is in capturing adequate abstractions accurately through mathematical models, requiring that the modeller captures the essential aspects of the protocol, while abstracting from certain details. Specifically, in our case, these are:

\begin{itemize}
	\item We abstracted from underlying web mechanisms such as OAuth 2.0, OpenID Connect, and communication protocols like TLS, considering them as abstract secure channels. These channels are natively handled by OFMC. For ProVerif, these abstract secure channels were compiled with a standard cryptographic implementation, based on symbolic asymmetric encryption, which provides the authentication and secrecy guarantees expected by such channels. This approach simplifies the verification task and mitigates the state explosion problem. As discussed in Section \ref{sub:methodology}, vertical protocol composition is sound under certain message disjointness conditions that are satisfied by the channels used in our model.
	\item We modelled functions that compute tokens as injective functions returning values modelled as nonces, so long as the input includes some sort of randomness. Intuitively, this ensures that the function returns a different value as long as the input is different at each call. This is a realistic assumption, since the protocol specifications dictate the use of random numbers and identifiers in many steps.
	\item While our model does not explicitly specify checks on tokens, the modelled channels provide sufficient conditions to satisfy the security goals of the protocols, such as authentication tokens. Our model shows that the application protocol could be implemented on simpler underlying technologies that provide basic communication guarantees (authentication and secrecy), rather than on the currently implemented web standards.  
	\item The \textit{AnBx} compiler automatically computes message checks on reception that recipients are expected to perform, including, for example, signature verification and matching of received values with their prior knowledge. This is essential for the construction of a robust implementation of the concrete system, as agents should be able to detect anomalies that diverges from expected behaviour, based on the knowledge acquired during the protocol execution so far. This applies also to the generation of the ProVerif model, where checks need to be declared explicitly, while in the AnBx model they are implicit. %
	\item The \textit{AnBx} compiler automatically generates a Java implementation of the protocol, as specified in the \textit{AnBx} model. This is not interoperable with the standard implementation, however, since Open Banking relies on specific web standards, as discussed before. Nevertheless, it provides evidence that other, potentially simpler, implementations of the protocol, may be possible.
	\item We verified $8$ security goals that are derived from the informal requirements stated in the documentation (see Section~\ref{sec:security}), and from our interpretation of the protocol security. However, we could have potentially missed other implicit goals. No explicit goals are given by the Open Banking standard itself.
	\item We are constrained by the security goals that the verification tools can handle, namely confidentiality and authenticity. Specifically, we could not model timing properties (e.g., token expiration), such as relay resistance~\cite{Drimer2007,Chothia2019}. These would require a notion of time, which is not available in \textit{AnBx}. %
\end{itemize}

Given the above considerations, there are scenarios where, if the assumptions in the model are not met in real-world implementation, the security of the system could be compromised. Potential scenarios include:

\begin{itemize}
	\item \textit{Poor Implementation or Misconfiguration of Underlying Protocols}: Misconfiguration or bugs in foundational protocols, such as TLS or OAuth 2.0, can lead to severe vulnerabilities. For example, if TLS is misconfigured (e.g., allowing outdated or weak cipher suites), attackers could perform man-in-the-middle (MITM) attacks, intercepting or modifying messages. Certificate validation issues may enable spoofing attacks, where an attacker impersonates an ASPSP or AISP. These scenarios invalidate the assumptions of confidentiality, integrity, and authentication that TLS is meant to provide. Furthermore, misconfiguration or a defective implementation of OAuth 2.0 may lead to issues, such as token leakage, token replay attacks, or insecure redirects.
	
	\item \textit{Session Management Issues}: Authentication sessions must be correctly handled and properly terminated. For example, if session tokens are not invalidated after logout or consent revocation, attackers may hijack active sessions. Moreover, if session handling lacks proper Cross-Site Request Forgery (CSRF) protection, an attacker could trick a logged-in user into performing unintended actions, such as authorising a malicious third-party application.
	
	\item \textit{Violation of Trust Assumptions}: Certain expectations exist regarding the behaviour of system agents. For example, a compromised AISP could misuse its access to PSU account data beyond the intended scope. Furthermore, while our model assumes that ASPSPs are trusted entities, insider threats at an ASPSP could manipulate token issuance to gain unauthorised access.
	
	\item \textit{Compromised Credentials}: The model relies on pre-shared secrets and abstract symbolic functions (e.g., \verb'fPSUSecret', \verb'fAISPSecret') for authentication. While these functions are assumed to be secure and private, their real-world implementation depends on specific mechanisms and applications defined at the organisational level. If such mechanisms are defective or vulnerable, the entire authentication system could be compromised. Additionally, in practice, phishing, brute-force attacks, or insecure storage may also lead to credential compromise.
	
	\item \textit{Access Control Issues}: Ineffective access control enforcement could allow attackers to exploit the system. For example, in the model, we assume that the PSU selects all accounts in the \verb'IntentAgreement', but in reality, a PSU might select only a subset of accounts. This discrepancy could lead to scenarios where, if the system fails to enforce access restrictions correctly, unauthorised access to unselected accounts could be granted.
\end{itemize}

It is also important to note that formal modelling and verification may not encompass all aspects of system security \cite{Kulik2022}, such as implementation flaws or side-channel attacks, which are better addressed by penetration testing. Penetration testing is convenient for simulating real-world attack scenarios, identifying configuration errors and implementation bugs, and leveraging extensive databases of known vulnerabilities \cite{Modesti2024EH}. Nonetheless, even with tools designed to test specific applications, such as OAuth 2.0 and Open ID Connect (e.g., \cite{Bisegna2020, Bisegna2020a}), penetration testing is limited to a specific point in time, cannot guarantee exhaustive coverage, and often focuses on known threats, rather than novel ones \cite{Bishop2007}. Therefore, penetration testing alone is insufficient, and many experts \cite{garavel20202020} recommend the inclusion of formal methods as part of a comprehensive and effective security assurance strategy.

In any case, for this research, penetration testing or automated tests were not possible, as it is not allowed by the terms and conditions of the sandbox we run experiments with (see~\ref{appendix:natwest-walk-through}). Moreover, such ethical hacking exercises are routinely performed by organisations, and our main research contribution is instead the formalisation and verification of the Account and Transaction API protocol. We used the tests presented in Section \ref{sec:sandbox-testing}, as a security sanity check and for the purpose of validating our model.

\section{Related Work}
\label{sec:relatedwork}

To the best of our knowledge, our work represents the first attempt to formally specify and automatically verify the Open Banking Account and Transaction API protocols, focusing on the application level and high-level security goals, while abstracting from the details of the underlying web technologies. To encode the model, we used the \textit{AnBx} language, a simple and intuitive notation, suitable for use by application developers and protocol designers.

There are several works that studied the underlying web technologies used by Open Banking. In particular, Fett et al. \cite{fett2019extensive} analysed the OpenID Financial Grade API (FAPI), which is a security profile based on OAuth 2.0 used by Open Banking, built on top of OIDC. Specifically, they examined aspects, such as authorisation, authentication, and session integrity. 

Additionally, Hosseyni et al. \cite{Hosseyni2024} conducted a formal security analysis of the FAPI 2.0 protocols, utilising an extended Web Infrastructure Model (WIM) to identify vulnerabilities. Authors uncovered several attacks, they collaborated with the FAPI Working Group to amend the protocols, and subsequently proved that the revised specifications satisfy the desired security properties. Interestingly, although they provide proofs to support their findings, the analysis was conducted on paper, as they state that ``creating a mechanized model of the Web is a challenge by itself and left for future work''. This leaves open the possibility that further issues might be uncovered by automated verification. In contrast, our analysis is automated, but our model abstracts from implementation details of the underlying web technologies. 

Furthermore, Fett et al. \cite{fett2016comprehensive} conducted a comprehensive formal security evaluation of OAuth 2.0, and other researchers have also investigated various security aspects of OAuth 2.0 (e.g. \cite{Philippaerts2022, Innocenti2023}) and OpenID Connect (e.g. \cite{Mainka2017, Navas2019}) protocols. Interestingly, Bisegna et al. \cite{Bisegna2020} developed a Burp Suite plug-in for automated OAuth/OIDC penetration testing, intended for the Micro-ID-Gym platform \cite{Bisegna2020a}. The plug-in conducts passive tests that analyse traffic for compliance and CSRF protection, and active tests that assess endpoint responses to altered parameters during the OAuth flow.

In contrast with the works cited so far, our contribution focuses on the protocol itself, which implements the financial transactional aspects of the standard at the application level, abstracting from the details of the underlying communication, cryptographic primitives or web mechanisms.

With a specific focus on Open Banking, Coste and Miclea \cite{coste2019api} tested the operations that trusted third parties (TTPs) can perform in relation to payment initiation service providers (PISPs) and account information service providers (AISPs). They evaluated PSD2 implementation compliance using testing tools such as Swagger and HP ALM for test execution and management, and validated error handling mechanisms to ensure API robustness and reliability. However, their paper did not explicitly state which specific implementation was tested.

Kellezi et al. \cite{kellezi2019towards,kellezi2021securing} evaluated the integration of a web application with Danish Nordea's Open Banking APIs (sandbox version), considering security threats in light of the OWASP Top 10 Web Application Security Risks. They focused on the implementation level, examining components like authentication and access control. However, unlike our work, they did not investigate the access authorisation flow, a core element of the Account and Transaction API protocol, nor did they explicitly state or test the security goals. However, like in our case, they were limited to studying a sandbox version of the API.

Behbehani et al. \cite{behbehani2022} conducted a threat modelling exercise proposing a STRIDE model to rank the security threats associated with the integration of third-party applications with the Open Banking API, using Bayesian analysis to predict the most likely exploitable attack paths. They also proposed a methodology \cite{behbehani2023open} for predicting anomalous access behaviour using machine learning techniques like random forests and linear kernel support vector machine (SVM) to compare accuracy. However, their dataset included only generic API calls and did not address specific Open Banking API calls.

Furthermore, Xu et al. \cite{Xu2020} proposed a blockchain-based data-sharing model for Open Banking called the Provenance-Provided Model (PPM). The system uses smart contracts to mediate between users and third-party services, incorporating changes at the data, smart contract, and application layers. The PPM model enables transparent authentication, privacy control, and auditable provenance.

Specifically regarding mobile technologies, many authors have proposed solutions to secure banking and payment services. A survey paper from 2021 \cite{Bojjagani2021} presents a systematic literature review of mobile payments, protocols, and security infrastructures, covering over 350 studies from 2000 to 2020.

Bojjagani and Sastry \cite{Bojjagani2017} propose a security framework for SMS-based mobile banking using elliptic curve cryptography, ensuring end-to-end secure communication between customers and banks. They validate its effectiveness through formal verification methods and security analysis. Similarly to our work, they conduct modelling and verification using two tools: AVISPA and Scyther in their case, and OFMC and ProVerif in ours. This approach involving multiple (compatible) formalisms with different strengths enables cross-validation and helps to overcome limitations of individual tools in handling specific cases and properties.

Aparicio et al. \cite{Aparicio2023} analyse vulnerabilities in SMS-based Two-Factor Authentication (2FA) using the SMS Retriever API, proposing a methodology to identify insecure server-side implementations in banking apps through app analysis. Their approach was validated by testing several Spanish banking services.

While these works focus on particular technical aspects of mobile technologies, the Open Banking protocol covered in this study is agnostic to certain implementation details, such as the client-side application, which can run independently on a web browser or mobile app. Additionally, mechanisms like MFA are delegated to the specific application implementation chosen by the financial institution and providers within the regulatory framework of PSD2.

Finally, Gounari et al. \cite{gounari2024harmonizing} have highlighted the lack of harmonisation and clarity on the technical security specifications necessary for the effective implementation of Open Banking, a significant challenge for fintech and banking organisations in understanding the full range of compliance requirements set forth by PSD2.

In light of work reviewed in this section, we believe that our formal modelling and automated verification of identified security goals, along with its abstraction from specific web technologies, provide valuable insights for various stakeholders, including financial institutions, regulatory bodies, third-party service providers, and security analysts. In particular, the formal documentation and verification of security goals are relevant given the significance and long-term impact of adopting a standard protocol for the efficiency, security, and interoperability of the financial sector \cite{Borgogno2021}.

\section{Conclusion}
\label{sec:conclusion}

This paper presents an \textit{AnBx} model of the Open Banking Account and Transaction API protocol and its automated verification against a set of security goals derived by our interpretation of the protocol documentation. Verification was carried out using two tools: OFMC~\cite{basin2005ofmc} and ProVerif~\cite{Blanchet2022}. OFMC enabled verification of all security goals (including strong authentication ones) for a single session, yet multiple sessions suffered from state-space explosion limitations. On the other hand, ProVerif successfully completed verification for an unbounded number of sessions in under four minutes, yet could not handle strong authentication for one of the eight security goals. 

Given that the Open Banking Account and Transaction protocol is a fundamental component of the systems deployed by numerous banks in the UK, with variations in other European countries in recent years \cite{Arner2022,DePascalis2024}, ensuring its correctness is paramount to maintaining trust in Open Banking services \cite{Kassab2022}. This will also help banks and the fintech sector prepare for the upcoming PSD3 and the Payment Systems Regulator (PSR), which will enforce more stringent security measures, including Strong Customer Authentication (SCA) (e.g., supporting authentication mechanisms beyond mobile app-based solutions) and enhanced fraud prevention and detection.
Additionally, the UK version of the standard represents the most developed in the market with a total of 14 billion API calls, compared to a combined 6.4 billion in Germany, Italy, France and Spain\cite{website:OB_stats-2024}. %

Overall, the Open Banking standard relies on current web technologies that may become outdated, as technological and security requirements evolve. Of greater concern are the broader security implications it imposes, as its security relies on the verification of numerous external dependencies. Our formalisation and analysis, abstracting from underlying web technologies, propose a path to reduce reliance on specific internet and web standards. It underscores the necessary assumptions and commitment specifications essential for any standard adopted within the Open Banking API. %

Moreover, it is worth noting that currently, the definition of the API is delegated to the national level. However, in light of the future adoption of PSD3 and PSR, the European Banking Authority (EBA) emphasises the importance of a common API standard across the EU. Although acknowledging that a single API standard would initially increase compliance costs, the EBA highlights significant long-term benefits, such as reducing the burden on Third-Party Providers (TPPs) to connect to ASPSPs' interfaces \cite{eba2022}. Additionally, a single unified API may enable a streamlined effort and enhance the impact of formal verification work similar to the one presented in this paper. 

Future work could proceed in several directions. One avenue is to formalise and verify other components of the standard or to model the protocol's state and specify transparent functions in formalisms like VDM-SL. This approach could potentially expose vulnerabilities related to low-level interactions with dependent technologies, such as OAuth2, OIDC, TLS, and others, as suggested by Kellezi et al. \cite{kellezi2021securing}.

\subsection*{Acknowledgement}
The authors thank the anonymous reviewers for their constructive feedback and express their gratitude to Oluwafemi Adesogbon, Callum Duncan and Musa Kabiri for engaging in constructive discussions during their student projects at Teesside University. Leo Freitas is grateful to the EPSRC EP/N023641/1 STRATA programme grant for partially financially supporting his work on this paper.

\bibliographystyle{elsarticle-num}
\bibliography{literature}

\newpage
\label{sec:appendix}
\appendix
\section{Abbreviations}
\label{sec:abbreviations}	
\renewcommand{\nomlabel}[1]{\textbf{#1}} 
\nomenclature[A]{TPP}{Third-Party Provider}
\nomenclature[A]{API}{Application Programming Interface}
\nomenclature[A]{AnB}{Alice and Bob Notation}
\nomenclature[A]{AnBx}{Extended Alice and Bob Notation}
\nomenclature[A]{AnBxC}{AnBx Compiler and Code Generator}
\nomenclature[A]{FI}{Financial Institution}
\nomenclature[A]{ASPSP}{Account Servicing Payment Service Provider}
\nomenclature[A]{AISP}{Account Information Service Provider}
\nomenclature[A]{PIS}{Payment Initiation Service}
\nomenclature[A]{PSU}{Payment Service User}
\nomenclature[A]{aspspR}{ASPSP -- Resource Server}
\nomenclature[A]{aspspA}{ASPSP -- Authorisation Server}
\nomenclature[A]{BPEL}{Business Process Execution Language}
\nomenclature[A]{OBIE}{Open Banking Implementation Entity}
\nomenclature[A]{CMA}{UK Competition and Markets Authority}
\nomenclature[A]{FCA}{UK Financial Conduct Authority}
\nomenclature[A]{PSD}{Payment Services Directive}
\nomenclature[A]{OBWG}{Open Banking Working Group}
\nomenclature[A]{JWS}{JSON Web Signature}
\nomenclature[A]{JWT}{JSON Web Token}
\nomenclature[A]{GDPR}{General Data Protection Regulation}
\nomenclature[A]{ATP}{Account and Transaction API Protocol}
\nomenclature[A]{TLS}{Transport Layer Security}
\nomenclature[A]{RS256}{RSA Signature with SHA-256}
\nomenclature[A]{PS256}{Probabilistic RSA Signature with SHA-256}
\nomenclature[A]{HS256}{HMAC Signature with SHA-256}
\nomenclature[A]{UUID}{Universally Unique Identifier}
\nomenclature[A]{SCA}{Strong Customer Authentication}
\nomenclature[A]{OIDC}{OpenID Connect}
\nomenclature[A]{FAPI}{Financial-grade API}
\nomenclature[A]{PPM}{Provenance-Provided Model}
\nomenclature[A]{PSR}{Payment Systems Regulation}
\nomenclature[A]{EBA}{European Bank Authority}
\nomenclature[A]{CSRF}{Cross-Site Request Forgery}
\nomenclature[A]{MITM}{Man-in-the-middle}
\nomenclature[A]{STRIDE}{Spoofing, Tampering, Repudiation, Information disclosure, Denial of service, Elevation of privilege}
\nomenclature[A]{2FA}{Two-Factor Authentication}
\nomenclature[A]{MFA}{Multi-Factor Authentication}
\nomenclature[A]{SVM}{Support Vector Machine}
\setlength{\nomitemsep}{-0.5\parsep}
\printnomenclature[1.8cm]

\newpage
\section{Formal Model (AnBx)}
\label{appendix:anb_model}
\begin{lstlisting}[language=AnBx,basicstyle=\ttfamily\scriptsize]
Protocol: ob_aisp_protocol AnBx

# Version: 14

# to generate ProVerif code, use anbxc <file> -expandbullets options
# only goal 6 requires -pvmovenewup for AnBx Compiler version 2025.04+, but the switch can be used for all goals

Types:

    # Four parties -- ASPSP = authorisation server + resource server
    Agent PSU, AISP, aspspA, aspspR;
    Certified PSU, AISP, aspspA, aspspR;
	
    # PSU Redirection URIs
    URI ASPSPAuthPSUEndpoint, AISPEndpoint;
    
    # Abstract away from specific intent details agreed on between PSU and AISP
    IntentNo IntentAgreement;
      
    # constants
    Number empty;
    # not really used, just to create the types, so that they can be used
    JWT nulljwt;			# JSON web Token
    AccountNumber nullaccount;

    # Access tokens are used to allow AISPs access to protected resources (including APIs), 
	# the tokens are only valid for a short duration (e.g. 30 minutes). 
    # Modelled as random values
    
    UUID AISPSecretNonce;
    UUID AccountsSecretNonce;
    
    # ---- ASPSP Authorisation Server Functionality ---- #

    # Return agreed on secret between AISP and authorisation server 
    # Params: AISP ID -- Output: the secret (private function)
    Function [Agent, Agent ->* Number] fAISPSecret;
    # Return agreed on secret between PSU and authorisation server
    # Params: PSU ID -- Output: the secret 
    Function [Agent, Agent ->* Number] fPSUSecret;
    # Generate a base level token for accessing consent creation API
    # Params: AISP ID, AISPSecretNonce -- Output: client token
    Function [Agent, UUID -> JWT] fClientCredToken;
    # Generate authorisation code exchangeable for an access token
    # Params: consent ID -- Output: authorisation code
    Function [JWT -> JWT] fAuthCode;
    # Create an access token for account data access
    # Params: authorisation code -- Output: authorisation code token
    Function [JWT -> JWT] fAuthCodeToken;
        
    # ---- ASPSP Resource Server Functionality ---- #
    
    # Create and return a consent resource
    # Params: client token, intent payload -- Output: consent ID
    Function [JWT, IntentNo -> JWT] fCreateIntent;
    # Retrieve the consent with the given identifier
    # Params: consent ID -- Output: consent resource
    Function [JWT -> IntentNo] fGetIntent; 
    # Return complete list of PSU accounts
    # Params: PSU ID -- Output: list of all PSU accounts with ASPSP
    Function [Agent -> AccountNumber] fFetchAccounts;
    # Change consent's status to authorised
    # Params: PSU ID, consent ID, selected PSU accounts -- Output: authorisation result
    Function [Agent, JWT, IntentNo -> JWT] fAuthoriseIntent;
    # Fetch PSU account(s) (request by AISP)
    # Params: authorisation code token, account identifier -- Output: account info.
    # Note: if account identifier = empty then it is a bulk call
    Function [JWT, UUID, Number -> AccountNumber] fAccountsEndpoint

Definitions:
    # AISP credentials to be authenticated by ASPSP
    AISPCredentials   : AISP, fAISPSecret(aspspA,AISP);
    
    # **RFC6749 4.4. Client Credentials Grant
    # https://datatracker.ietf.org/doc/html/rfc6749#section-4.4
    # For asking for a client token
    ClientTokenReq    : AISPCredentials;
    
    # For returning a client token
    # Token generated through the Client Credential Grant (refer to it as client token)
    # The AISPSecretNonce param allow to generate a different client token for every session
    ClientToken       : fClientCredToken(AISP,AISPSecretNonce);
    ClientTokenRes    : AISPSecretNonce,ClientToken;				# the nonce must be sent in the response, x-fapi-interaction-id: A UUID that provides a unique reference to be created for each request/response.

    # **OB Account Access Consents v3.1.11
    # For asking for the creation of an intent
    # Assume that the risk data is included with the intent agreement
    IntentReq         : ClientToken, IntentAgreement;
    # For returning for the created intent
    # Treat the intent as an intent (AccountRequestId) or a complete intent object (abstraction)
    Intent  		  :	fCreateIntent(ClientToken, IntentAgreement);
    
    #  ------- Consent Authorisation (Authorisation Code Grant) -------
    # **RFC6749 4.1. Authorisation Code Grant + OB Account and Transaction API Sequence Diagram (Step 3) 
    # Redirection command given by AISP to PSU to redirect to ASPSP and authorise consent
    RedirectToASPSPCmd: ASPSPAuthPSUEndpoint, AISPEndpoint, Intent;
    # PSU Credentials to be authenticated by ASPSP
    PSUCredentials    : PSU, fPSUSecret(PSU,aspspA);
    # Payload sent by PSU to ASPSP authorisation server to authenticate themselves and initiate consent authorisation
    InitAuthIntentReq : RedirectToASPSPCmd, PSUCredentials;
    
    # For PSU consent review and account selection
    RetrieveIntentReq : Intent, PSU;
    RetrieveIntentRes : fGetIntent(Intent), fFetchAccounts(PSU);
    
    # For updating consent resource's state
    
    SelectedAccounts: IntentAgreement;			# We assume here that the PSU will select all accounts in the IntentAgreement
    
    AuthoriseIntentReq: PSU, Intent, SelectedAccounts;
    AuthoriseIntentRes: fAuthoriseIntent(PSU, Intent, SelectedAccounts);
    
    # For generating authorisation code
    AuthorisationCode : fAuthCode(Intent);
    # Redirection command given by ASPSP to PSU to redirect to AISP
    RedirectToAISPCmd : AISPEndpoint, AuthorisationCode;
    #  ----------------------------------------------------------------
    
    # **RFC6749 4.1. Authorisation Code Grant
    # For asking for an authorisation token
    AuthTokenReq      : AISPCredentials, AuthorisationCode;
    # For returning an authorisation token
    AuthToken         : fAuthCodeToken(AuthorisationCode);                                                        
    
    # **OB Accounts v3.1.1
    # For asking for the PSU accounts list
    AccountsReq       : AuthToken;  
    
    # For returning the PSU accounts list
    Accounts          : fAccountsEndpoint(AuthToken, AccountsSecretNonce, empty);
    
Knowledge:
    PSU      : PSU, AISP, aspspA, aspspR;
    AISP     : AISP, PSU, aspspA, aspspR;

    aspspA: PSU, AISP, aspspA, aspspR, fClientCredToken, fAuthCode, fAuthCodeToken;
    aspspR : PSU, AISP, aspspA, aspspR, fCreateIntent, fGetIntent, fFetchAccounts, fAuthoriseIntent,fAccountsEndpoint;
    
    # FIX agents share authentication credentials
    aspspA,PSU share fPSUSecret(PSU,aspspA);
    aspspA,AISP share fAISPSecret(aspspA,AISP);

    # Impose condition that a PSU cannot also be the AISP or any of the ASPSP's servers
    # Impose condition that AISP cannot be ASPSP due to unrealistic conditions if allowed in AnB 
    
    where PSU!=AISP,PSU!=aspspA,PSU!=aspspR,AISP!=aspspA,AISP!=aspspR
    
Actions:
    #   ----- Step 1: request account information -----
    # The flow begins with a PSU consenting to allow an AISP to access account information data. 
   
    # https://standards.openbanking.org.uk/customer-experience-guidelines/account-information-services/account-information-consent/latest/
    # the PSU indicates the resources and the ASPSP (bank)
    PSU *->* AISP       : IntentAgreement, PSU, aspspA    #A1.1
    
    #   ----- Step 2: setup account access consent ----
    # the AISP authenticate with the aspspA using pre-shared credentials AISPCredentials aka ClientTokenReq
    # https://datatracker.ietf.org/doc/html/rfc6749#section-4.4
    
    #/account-access-consents
    # https://openbankinguk.github.io/read-write-api-site3/v3.1.11/resources-and-data-models/aisp/account-access-consents.html#post-account-access-consents
    AISP *->* aspspA    : ClientTokenReq     #A2.1  initiate client credential grant Get account/transaction information POST Client Credential Grant ) account scope
    aspspA *->* AISP    : ClientTokenRes     #A2.2  access-token Response: { access_token=" ", token_type="Bearer", expires_in=3600 }

    
    # This resource indicates the consent that the AISP claims it has been given by the PSU to retrieve account and transaction information.
    AISP *->* aspspR     : IntentReq #A2.3  post /account-access-consents Image 6
    aspspR *->* AISP     : Intent    #A2.4  HTTP 201 (Created), ConsentID Response { ConsentID = " ",  other data: permissions, Risk... }
    
    
    #   ----- Step 3: authorise consent -----
    #   (using authorisation code grant)
    #       ----- Step 3.1 -----
    AISP *->* PSU       : RedirectToASPSPCmd #A3.1.1 GET /auth-code-url/.../scope={requestscope}+++ HTTP 302 (Found), Redirect (ConsentId) - get URL copy/paste in browser
    PSU *->* aspspA     : InitAuthIntentReq  #A3.1.2 Follow redirect (ConsentId)	authenticate (PSU user name and password)	

    #       ----- Step 3.2 -----
    aspspA *->* aspspR: RetrieveIntentReq    #A3.2.1 request authentication		
    aspspR *->* aspspA: RetrieveIntentRes    #A3.2.2
    aspspA *->* PSU     : RetrieveIntentRes  #A3.2.3
    PSU *->* aspspA     : SelectedAccounts   #A3.2.4 select accounts
    aspspA *->* aspspR: AuthoriseIntentReq   #A3.2.5 Exchange authorization-code for access token GET auth-code-url
    aspspR *->* aspspA: AuthoriseIntentRes   #A3.2.6
    #       ----- Step 3.3 -----    
    aspspA *->* PSU     : RedirectToAISPCmd  #A3.3.1 HTTP 302 (Found), Redirect (authorization-code) Response {authcode from url ...}
    PSU *->* AISP       : RedirectToAISPCmd  #A3.3.2 Follow redirect (authorization-code)
    AISP *->* aspspA    : AuthTokenReq       #A3.3.3 Poll at /token using auth-req-id (auth code grant) code=authtoken)
    aspspA *->* AISP    : AuthToken          #A3.3.4 access-token	
    
    #   ----- Step 4: request data ------
    AISP *->* aspspR      : AccountsReq      #A4.1 GET /accounts/{AccountID}/transactions
    aspspR *->* AISP      : Accounts, PSU  	 #A4.2 /HTTP 200, List of transactions (AccountRes)
  
Goals:
    
    # all goals OK with OFMC 1 session, ProVerif unbounded, but G4 only for weak auth
	# only goal 6 requires -pvmovenewup for AnBx Compiler version 2025.04+, but the switch can be used for all goals
    
    fAISPSecret(aspspA,AISP) secret between AISP, aspspA  #G1 PV OK (mutual)
    fPSUSecret(PSU,aspspA) secret between PSU, aspspA  #G2 PV OK (mutual)
    ClientToken, AuthToken secret between AISP, aspspA, aspspR  #G3PV OK (mutual)

    PSU weakly authenticates aspspA on fGetIntent(Intent)  #G4 PV OK (mutual)
    # PSU authenticates aspspA on fGetIntent(Intent)  #G4 PV loops
    
    aspspR weakly authenticates PSU on SelectedAccounts  #G5 PV OK (mutual)
    
    PSU weakly authenticates AISP on ASPSPAuthPSUEndpoint, AISPEndpoint  #G6 PV OK (mutual)
    
    Accounts secret between AISP, aspspR  #G7 PV OK (mutual)
    
    AISP weakly authenticates aspspR on Accounts  #G8 PV OK (mutual)
\end{lstlisting}

\newpage
\section{NatWest Open Banking Walk-through}
\label{appendix:natwest-walk-through}
This appendix provides an example of protocol execution, as outlined in Figure~\ref{fig:anb_aisp_seq_diagram} in Section~\ref{sec:implementation}.

\subsection{ASPSP Data Capabilities and Account Information Request}

The PSU initiates account information sharing with the AISP by sending an intent agreement (\textit{A1.1}). This agreement defines the scope of shared data, specifying permissions and access restrictions. Its nature is crucial, as it determines the level of data access granted to the AISP.
To construct the intent agreement, the AISP must understand the ASPSP's authorisation server capabilities. This is achieved by querying the server to determine the scope and accessibility of available data.

The AISP sends a GET request to \smallurl{https://api.sandbox.natwest.com/.well-known/openid-configuration} to discover the ASPSP's data capabilities. This request requires no specific parameters, headers, or body beyond the standard GET request. The server responds with a successful  HTTP status code (\textit{200 OK}) along with a JSON object detailing NatWest's sandbox data capabilities (Figure~\ref{fig:AccessTo}).

The response includes information such as the API version, aligning client-server expectations and ensuring support for new features. It also provides the issuer URL for verifying access token authenticity, as well as authorisation and token endpoints, supported scopes, grant types, authentication methods, and signing algorithms. The authorisation and token endpoints enable secure OAuth 2.0 flows, while the \textit{jwks\_uri} field points to the JSON Web Key Set, essential for maintaining token integrity.

\begin{figure}[th]
  \centering
  \begin{lstlisting}[language=json]
{
    "version": "3.0",
    "issuer": "https://api.sandbox.natwest.com",
    "authorisation_endpoint": "https://api.sandbox.natwest.com/authorize",
    "token_endpoint": "https://ob.sandbox.natwest.com/token",
    "jwks_uri": "https://keystore.openbankingtest.org.uk/0015800000jfwxXAAQ/0015800000jfwxXAAQ.jwks",
    "registration_endpoint": "https://ob.sandbox.natwest.com/register",
    "scopes_supported": [
        "openid","payments","accounts","fundsconfirmations","profile"
    ],
    "claims_supported": ["aud","exp","iat","iss","acr","openbanking_intent_id","sub"
    ],
    "acr_values_supported": [
        "urn:openbanking:psd2:ca"
    ],
    "response_types_supported": [
        "code id_token"
    ],
    "response_modes_supported": [
        "fragment"
    ],
    "grant_types_supported": [
        "authorisation_code",
        "refresh_token",
        "client_credentials"
    ],
    "subject_types_supported": [
        "public"
    ],    "id_token_signing_alg_values_supported": [
        "PS256"
    ],    "token_endpoint_auth_methods_supported": [
        "tls_client_auth",
        "private_key_jwt"
    ],
    "token_endpoint_auth_signing_alg_values_supported": [
        "PS256"
    ],
    "claim_types_supported": [
        "normal"
    ],
    "claims_parameter_supported": true,
    "request_parameter_supported": true,
    "request_uri_parameter_supported": false,
    "request_object_signing_alg_values_supported": [
        "PS256"
    ],
    "request_object_encryption_alg_values_supported": [],
    "request_object_encryption_enc_values_supported": [],
    "tls_client_certificate_bound_access_tokens": true
}
\end{lstlisting}
  \caption{Successful response from ASPSP sandbox about its data capabilities}\label{fig:AccessTo}
\end{figure}

\subsection{Setup Account Request}

In the second step, the AISP requests a client token from the ASPSP Authorisation Server (aspspA) to initiate the creation of a consent resource. A POST request is sent to \smallurl{https://ob.sandbox.natwest.com/token} with parameters \texttt{grant\_type}, \texttt{client\_id}, \texttt{client\_secret} and \texttt{scope} in the request body, and the Content-Type header set to \texttt{application/x-www-form-urlencoded} (\textit{A2.1}). These parameters are required for authentication.

Once authenticated, the ASPSP Authorisation Server issues the client token (\textit{A2.2}), as shown in Figure~\ref{fig:Access}, enabling the AISP to request the creation of a consent resource from the ASPSP Resource Server (aspspR) (\textit{A2.3}). The aspspR then generates a consent resource based on the agreed intent between the AISP and PSU (\textit{A2.4}) and returns an identifier to the AISP for future reference.

\begin{figure}[H]
  \centering
  \begin{lstlisting}[language=json]
{
    "access_token": "eyJhbGciOiJSUzI1NiIsInR5cCI6IkpXVCJ9.eyJhcHAiOiJRdWR1cyBBcHAiLCJvcmciOiJsaW5rZmllbGRjYXJlLmNvbSIsImlzcyI6Imh0dHBzOi8vYXBpLnNhbmRib3gubmF0d2VzdC5jb20iLCJ0b2tlbl90eXBlIjoiQUNDRVNTX1RPS0VOIiwiZXh0ZXJuYWxfY2xpZW50X2lkIjoiZUpzdWhGZHRLNmFKRXRwVk8wOGN5d29BWHZYMWljQ1FKbVZ6M1FvdTNraz0iLCJjbGllbnRfaWQiOiI3MTcwZjlmMC1jNDk2LTRiNGItYjZlNy01MGMyMDVmZTFjMDkiLCJtYXhfYWdlIjo4NjQwMCwiYXVkIjoiNzE3MGY5ZjAtYzQ5Ni00YjRiLWI2ZTctNTBjMjA1ZmUxYzA5Iiwic2NvcGUiOiJhY2NvdW50cyIsImV4cCI6MTcxMzg2MzQ5NiwiaWF0IjoxNzEzODYyODk2LCJqdGkiOiJlNzlkOTAzMi1hOTJlLTQxNWUtODg0NS03Mzc4N2UwYWJjMjEiLCJ0ZW5hbnQiOiJOYXRXZXN0In0.Fqt9BRADyPE8VIw1DgzPFjC6ip20VZaEx6ICTO2wZDyqX-4t1F05v156DPzqJyyJtBbhV8f7NKeFQ-Ma_lurT-ijjB-zn3AxiM6TbXdXW0h2tHebyJDhlclS5vACZsEYdZGx7bemEtruCKPTSuH9k3shTznnGvan6eG2LjsxKO0MiB1eK0l5cagDKvGy9V4ojba4Hnu1ZyWOz_ga7YL17qSdYr99yW0Ga49LkZOYqAiWOhfGIMhyD0FBZSdCT2_-mpjRc5pRHcPCQB83EN7sEmfAc4-6qCWhBrNNCLZJXhJ8y8jGkVClBa6ForcpJ4r_DrXSx5QCuyVAQm7kKD8HWw",
    "token_type": "Bearer",
    "expires_in": 600,
    "scope": "accounts"
}
\end{lstlisting}
  \caption{Successful response to access token request (Action \textit{A2.2})}\label{fig:Access}
\end{figure}

The response obtained from (\textit{A2.2}) is shown in Figure \ref{fig:Access}. It includes an access token used for authentication, a token type indicating it is a bearer token, an expiration time of 600 seconds, and a scope specifying the permissions granted to the token. The decoded client token response and its security implications are discussed in Section \ref{sub:access-token}.

\subsection{Authorise Consent}
Authorisation is carried out in three stages: initiation, review, and finalisation. The process begins with the AISP initiating a POST request for authorisation to \smallurl{https://ob.sandbox.natwest.com/open-banking/v3.1/aisp/account-access-consents} \textit{(A3.1)}, with specific parameters in the request header (\textit{Authorisation = generated access token} and \textit{Content\_type set to application/json}). The JSON payload in Figure \ref{fig:permission} is included in the request body. It specifies the types of permissions being requested by the AISP.

\begin{figure}[H]
  \centering
  \begin{lstlisting}[language=json]
{
  "Data": {
    "Permissions": [
      "ReadAccountsDetail",
      "ReadBalances",
      "ReadTransactionsCredits",
      "ReadTransactionsDebits",
      "ReadTransactionsDetail"
    ]
  },
  "Risk": {}
}
\end{lstlisting}
  \caption{JSON payload for requesting account data permissions  (Action \textit{3.1})}\label{fig:permission}
\end{figure}

\begin{figure}[H]
  \centering
  \begin{lstlisting}[language=json]
{
    "Data": {
        "ConsentId": "adf5ae88-7382-4983-9a62-eb4cb809c593",
        "CreationDateTime": "2024-04-23T09:13:24.358Z",
        "Status": "AwaitingAuthorisation",
        "StatusUpdateDateTime": "2024-04-23T09:13:24.366Z",
        "Permissions": [
            "ReadAccountsDetail",
            "ReadBalances",
            "ReadTransactionsCredits",
            "ReadTransactionsDebits",
            "ReadTransactionsDetail"
        ]
    },
    "Risk": {},
    "Links": {
        "Self": "https://ob.sandbox.natwest.com/open-banking/v3.1/aisp/account-access-consents/adf5ae88-7382-4983-9a62-eb4cb809c593"
    },
    "Meta": {
        "TotalPages": 1
    }
}
\end{lstlisting}
  \caption{Successful response to initiating authorisation consent (Action \textit{3.1})}\label{fig:Action3.1}
\end{figure}

The JSON response in Figure \ref{fig:Action3.1} encapsulates an ongoing consent authorisation process between the AISP and PSU for accessing account information. It includes key elements such as \texttt{ConsentId}, \texttt{CreationDateTime}, and \texttt{Status}, which are essential for tracking and managing consents throughout their lifecycle. The \texttt{Status} field indicates the current state of the consent, typically \textit{AwaitingAuthorisation} until the customer approves access. The permissions specify the data accessible to the application, ensuring tailored access based on user preferences. While the \texttt{Risk} section may be empty if no concerns are identified, the \texttt{Links} section provides resources for direct interaction with the consent. The \texttt{Meta} section offers additional metadata, which aids in handling pagination or retrieving a comprehensive view of consents. Unique identifiers and creation timestamps reflect each interaction's evolution. These elements highlight the importance of consent tracking and management for compliance, security, and user control.

Following the creation of the consent resource (Figure \ref{fig:Action3.1}), the AISP instructs the PSU to redirect to aspspA and request consent authorisation. The PSU then interacts with aspspA to authorise the consent (\textit{A3.1.2}).

\begin{figure}[H]
  \centering
  \begin{lstlisting}[language=json]
{
    "redirectUri": "https://webbanking.sandbox.natwest.com/?domainName=linkfieldcare.com#/#code=6e945d68-77c3-4e37-9086-04bed7a8d9ee&id_token=eyJhbGciOiJQUzI1NiIsInR5cCI6IkpXVCIsImtpZCI6IndGWmxENDVodjQ1Z09zcThEYWMzbHh3NUZwQSJ9.eyJzdWIiOiJhZGY1YWU4OC03MzgyLTQ5ODMtOWE2Mi1lYjRjYjgwOWM1OTMiLCJhY3IiOiJ1cm46b3BlbmJhbmtpbmc6cHNkMjpzY2EiLCJhdWQiOiJlSnN1aEZkdEs2YUpFdHBWTzA4Y3l3b0FYdlgxaWNDUUptVnozUW91M2trPSIsImNfaGFzaCI6ImJPcTBFdDVUczNBTHlfcGFFVi02U3ciLCJvcGVuYmFua2luZ19pbnRlbnRfaWQiOiJhZGY1YWU4OC03MzgyLTQ5ODMtOWE2Mi1lYjRjYjgwOWM1OTMiLCJzX2hhc2giOiJ0ZFFFWEQ5R2I2a2Y0c3hxdm5raktnIiwiYXV0aF90aW1lIjoxNzEzODY0NTcwLCJpc3MiOiJodHRwczovL2FwaS5zYW5kYm94Lm5hdHdlc3QuY29tIiwiZXhwIjoxNzQ1NDAwNTcwLCJ0b2tlbl90eXBlIjoiSURfVE9LRU4iLCJpYXQiOjE3MTM4NjQ1NzAsImp0aSI6ImZhMGFhZjRmLTY0YTctNDdiYi1hODQ2LWU2MGFkMzViMmJmNSJ9.UUd9nq4mZF94EP3LbKuAUD-L3FQcVy7emlgcvYn6rk-nXzdaywZWcank2e0-OIJPuypDdZXOaQ5HCt8yVGbatAksZ9am127CTQzvzz-Q1yFTcmpTI56DqciYCXlaDVkSdo3oDHHOQWupotbFmya4ltOhSfkwEgnJuRsI1Wlp3VB7sugP8yikhPUXpTW9VQkJqC4XxMuQ9QpBt-YrIsBB_BC5etKen_EE_zJURxqzadaFaW89w-nVFNjKUssM0ARhP22kv7u9HH-r_1-8OK5HRoyW8gZ93E6SEXstsCigDU6ZysGaoLY_s-vCyUuMKjl2DyoMIVW-jjMyLWkDJXDeuQ&state=ABC"
}
\end{lstlisting}
  \caption{Successful response to AISP redirectUri to PSU (Action \textit{A3.1.1})}\label{fig:A}
\end{figure}
To obtain the response shown in Figure \ref{fig:A}, which contains the \texttt{redirectUri}, a GET request was made to \smallurl{https://api.sandbox.natwest.com/authorize?client_id={{clientId}}&response_type=code\%20id_token&scope=openid\%20accounts&redirect_uri={{encodedRedirectUrl}}&state=ABC&request={{consentId}}&authorization_mode=AUTO_POSTMAN&authorization_username={{psuUsername}}} with all requested parameters input manually or configured and saved in the Postman environment for dynamic use. After sending the request, a \textit{200 OK} response was returned, along with the \texttt{redirectUri} link for authorisation, as expected.

\begin{figure}[H]
    \centering
    \includegraphics[width=1\linewidth]{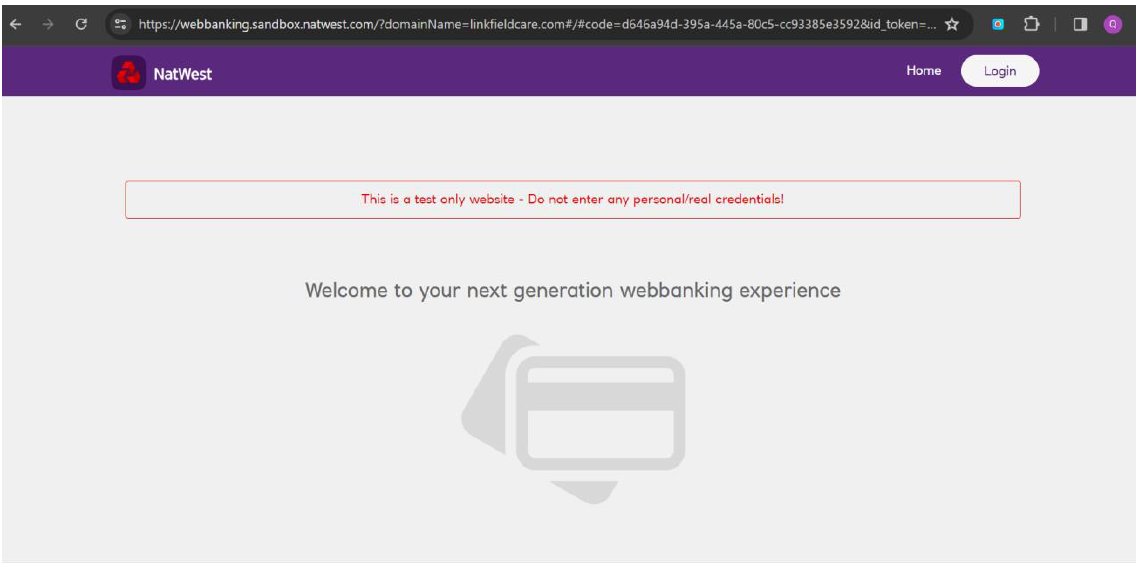}
    \caption{The page redirected to for authorisation after clicking on the redirectUri as shown in the Figure \ref{fig:A} (Action \textit{A3.2.3})}
    \label{fig:A3.2.3}
\end{figure}

After the PSU receives the redirect request and authenticates, upon successful authentication, aspspA requests the necessary resources from aspspR for the PSU to review before authorising the consent (\textit{A3.2.1}). aspspR provides these resources to aspspA (\textit{A3.2.2}), which then presents them to the PSU for review and consent (\textit{A3.2.3}, Figure \ref{fig:A3.2.3}). Following PSU authorisation, the PSU informs aspspA about the associated accounts linked to the consent (\textit{A3.2.4}), and subsequently, aspspA communicates this authorisation to aspspR (\textit{A3.2.5}). aspspR updates the consent resource's state and notifies aspspA of success (\textit{A3.2.6}). In the sandbox, we are unable to observe communication between aspspA and aspspR for actions \textit{A3.2.1}, \textit{A3.2.2}, \textit{A3.2.5}, and \textit{A3.2.6}.

After clicking the \texttt{redirectUri} in Figure \ref{fig:A} and authorising the consent resource, aspspA instructs the PSU to redirect back to the AISP to request an access token (\textit{A3.3.1}). The PSU provides the necessary information to the AISP (\textit{A3.3.2}), which then requests an authorisation token for the PSU account data from aspspA (\textit{A3.3.3}). Part of the data used for the request is obtained after consent has been granted. The access token is then sent to the AISP by aspspA (\textit{A3.3.4}).

\begin{figure}[t]
  \centering
  \begin{lstlisting}[language=json]
{
    "access_token": "eyJhbGciOiJSUzI1NiIsInR5cCI6IkpXVCJ9.eyJhcHAiOiJRdWR1cyBBcHAiLCJvcmciOiJsaW5rZmllbGRjYXJlLmNvbSIsImlzcyI6Imh0dHBzOi8vYXBpLnNhbmRib3gubmF0d2VzdC5jb20iLCJ0b2tlbl90eXBlIjoiQUNDRVNTX1RPS0VOIiwiZXh0ZXJuYWxfY2xpZW50X2lkIjoiZUpzdWhGZHRLNmFKRXRwVk8wOGN5d29BWHZYMWljQ1FKbVZ6M1FvdTNraz0iLCJjbGllbnRfaWQiOiI3MTcwZjlmMC1jNDk2LTRiNGItYjZlNy01MGMyMDVmZTFjMDkiLCJtYXhfYWdlIjo4NjQwMCwiYXVkIjoiNzE3MGY5ZjAtYzQ5Ni00YjRiLWI2ZTctNTBjMjA1ZmUxYzA5IiwidXNlcl9pZCI6IjEyMzQ1Njc4OTAxMkBsaW5rZmllbGRjYXJlLmNvbSIsImdyYW50X2lkIjoiMTgzOWY1YzEtOGEwMS00NmJhLWJlMmMtYTI3NmJjNWE3ZGZlIiwic2NvcGUiOiJhY2NvdW50cyBvcGVuaWQiLCJjb25zZW50X3JlZmVyZW5jZSI6IjJjYmY1ZTlkLTg0YmItNDFkMS05NmUwLWM5NzM3ZjdjMGNkYSIsImV4cCI6MTcxMzg4NzMyMCwiaWF0IjoxNzEzODg2NzIwLCJqdGkiOiJjNTBkYTBiYS0wYWVjLTQ2OWItOWI4My02ZWI3ZDZhNDAxNzYiLCJ0ZW5hbnQiOiJOYXRXZXN0In0.L07iQO57t2nxzo6mtEwe1L9f-NFgx5XsrEFnxWFCSuTN_mL6dn-x5Vr-usMgqcYKBZUcOk6S72nNYiONzgEl4u5eQ0cLwLGqZLbr9kD4GydUfJreTZH1h-Yz0a0bBQhftBlsJqwaz7BMsZOQJeOHooRiWq_nwV9Z2kP2oqAVOAQ-HObE7brwNeiReQ729iZl0Ocf8O55Ohmb6JfkkV4axpdZYaAcr6BKyZxe9SSNyVBfu8-kkxY80oZqTj-yvx2p6DX-tJ89vshhTQWcKuo9prVP_F9-_uy8t3QrKQgMYNrYuXnd0TgVz7xrPl2jTBc7TMdjYt1gtx06Ed9BdyNMmw",
    "refresh_token": "eyJhbGciOiJSUzI1NiIsInR5cCI6IkpXVCJ9.eyJhcHAiOiJRdWR1cyBBcHAiLCJvcmciOiJsaW5rZmllbGRjYXJlLmNvbSIsImlzcyI6Imh0dHBzOi8vYXBpLnNhbmRib3gubmF0d2VzdC5jb20iLCJ0b2tlbl90eXBlIjoiUkVGUkVTSF9UT0tFTiIsImV4dGVybmFsX2NsaWVudF9pZCI6ImVKc3VoRmR0SzZhSkV0cFZPMDhjeXdvQVh2WDFpY0NRSm1WejNRb3Uza2s9IiwiY2xpZW50X2lkIjoiNzE3MGY5ZjAtYzQ5Ni00YjRiLWI2ZTctNTBjMjA1ZmUxYzA5IiwibWF4X2FnZSI6ODY0MDAsImF1ZCI6IjcxNzBmOWYwLWM0OTYtNGI0Yi1iNmU3LTUwYzIwNWZlMWMwOSIsInVzZXJfaWQiOiIxMjM0NTY3ODkwMTJAbGlua2ZpZWxkY2FyZS5jb20iLCJncmFudF9pZCI6IjE4MzlmNWMxLThhMDEtNDZiYS1iZTJjLWEyNzZiYzVhN2RmZSIsInNjb3BlIjoiYWNjb3VudHMgb3BlbmlkIiwiY29uc2VudF9yZWZlcmVuY2UiOiIyY2JmNWU5ZC04NGJiLTQxZDEtOTZlMC1jOTczN2Y3YzBjZGEiLCJpYXQiOjE3MTM4ODY3MjAsImp0aSI6Ijc0NWY4Y2E3LWRjNDYtNGFhYy1hMWJkLWUzY2YwODc0MDFkMyIsInRlbmFudCI6Ik5hdFdlc3QifQ.ZUhmXO5A03p5uFD89uayc6dgLI-OCmziWOD5-kDX2jf68I53pZwB5G3KM8qU33GpyFPPlTn27OhS37305_lWCNZhEtCnrw3siTyaPnZhBWHQfVg_sGZV6_H38X54OOKVe0fJhE0xgiiHQD8_foQJb6JGvoq_ztayJrJNhpQAmWqAaOC5o55xTWU0z28YdJmrYNPA4ElQEG-aqf3RzMwZewzRS5ZsdCIWqECpN-D3wm_iWiMEhfNuPk6HvIiemL_GM1c3ohOZsFiTzt-Rq_pZvrCoIWcAgxRMa85Z4Vg28zHNctka5D7Fgy7RXp-AX6B7jVkTI4WSFta-tRRd_k3new",
    "id_token": "eyJhbGciOiJQUzI1NiIsInR5cCI6IkpXVCIsImtpZCI6IndGWmxENDVodjQ1Z09zcThEYWMzbHh3NUZwQSJ9.eyJzdWIiOiJjOWY1ODE5ZC04ZDE5LTQ1NWYtYTJmYy1lMDVkZTVlNWZkYzkiLCJhY3IiOiJ1cm46b3BlbmJhbmtpbmc6cHNkMjpzY2EiLCJhdWQiOiJlSnN1aEZkdEs2YUpFdHBWTzA4Y3l3b0FYdlgxaWNDUUptVnozUW91M2trPSIsImNfaGFzaCI6InVkOEdyalJsb1FIV2JKYXVfQnhlb0EiLCJvcGVuYmFua2luZ19pbnRlbnRfaWQiOiJjOWY1ODE5ZC04ZDE5LTQ1NWYtYTJmYy1lMDVkZTVlNWZkYzkiLCJzX2hhc2giOiJ0ZFFFWEQ5R2I2a2Y0c3hxdm5raktnIiwiYXV0aF90aW1lIjoxNzEzODg2NzE1LCJpc3MiOiJodHRwczovL2FwaS5zYW5kYm94Lm5hdHdlc3QuY29tIiwiZXhwIjoxNzQ1NDIyNzE1LCJ0b2tlbl90eXBlIjoiSURfVE9LRU4iLCJpYXQiOjE3MTM4ODY3MTUsImp0aSI6ImYxM2JlMTg0LTM2ZTAtNDgxYi1hNjIzLThmNTMwOWE5NTE4YyJ9.XwhjBqnyUgSob4QzQqrfjTF27-SJPeJNVQGwAoPTW6EdCOePQP4xG1oOGAQTykgmfkpDnP6jlGaXw3EZb8AqHjpx2i2L-JYLuZoeUmDUm_PYDut53E5f4UAj2TbsIhOv8o4BIdOY6Sjn9KJ17hfyfB4iCKqu0Ll7KV_ASmiyP1dBh4PFB8YI7FpXnb-iDFKjzCxMjbanL0aD1KVMadl5CbOYpHk75p2s3pXLbc43g_Hh7Mb9tuVougiruMXyEu6hSzAy0JsVNdhThOE8EdEA-mc3IqKYDwd23k2IDZvMEs5pAurJSSZBqnyXIuA81P6mYjp8lXfMc47QizzArTcA_g",
    "token_type": "Bearer",
    "expires_in": 600,
    "scope": "accounts openid"
}
\end{lstlisting}
  \caption{Successful response to authorisation token request (Action \textit{A3.3.4})}\label{fig:AccessTokenA3.3.4}
\end{figure}

To obtain the response shown in Figure \ref{fig:AccessTokenA3.3.4}, which includes an \textit{access\_token}, \textit{refresh\_token}, and \textit{id\_token} for authorisation, we initiated a token request to \smallurl{https://ob.sandbox.natwest.com/token} via a POST call. The request body included parameters such as \textit{grant\_type}, \textit{(authorisation\_code)} from the previous authentication step, \textit{redirect\_uri}, \textit{client\_id}, and \textit{client\_secret}. Upon successful validation, a \textit{200 OK} response was returned, indicating the request’s success, along with the access, refresh, and identity tokens, as well as token metadata. This response enables subsequent secure API interactions on behalf of the authenticated user.

\subsection{Request Data}
Finally, upon receiving the request for the authorisation token, aspspA validates the request and the provided authorisation code. If the code is valid and corresponds to an authorised consent, the AISP sends a GET request to \smallurl{https://ob.sandbox.natwest.com/open-banking/v3.1/aisp/accounts} with an authentication header containing the access token obtained after consent authorisation (\textit{A4.1}). In response, aspspR verifies the token’s validity and retrieves the permitted account data for the AISP (\textit{A4.2}) for further processing or analysis (Figure \ref{fig:AccessTokenA4.2]}).

\begin{figure}[H]
  \centering
  \begin{lstlisting}[language=json]
{
    "Data": {
        "Account": [
            {
                "AccountId": "31aa4a83-67bd-4e25-9e2f-7498303b076d",
                "Currency": "GBP",
                "AccountType": "Personal",
                "AccountSubType": "CurrentAccount",
                "Description": "Personal",
                "Nickname": "Sydney Beard",
                "Account": [
                    {
                        "SchemeName": "UK.OBIE.SortCodeAccountNumber",
                        "Identification": "50000012345601",
                        "Name": "Sydney Beard"
                    }
             {
                "AccountId": "7ee7091c-a5a2-4b78-b55e-f2c502060798",
                "Currency": "GBP",
                "AccountType": "Personal",
                "AccountSubType": "Savings",
                "Description": "Personal",
                "Nickname": "Sydney Beard",
                "Account": [
                    {
                        "SchemeName": "UK.OBIE.SortCodeAccountNumber",
                        "Identification": "50000012345602",
                        "Name": "Sydney Beard"
                 ]
    },
    "Links": {
        "Self": "https://ob.sandbox.natwest.com/open-banking/v3.1/aisp/accounts"
    },
    "Meta": {
        "TotalPages": 1
    }
}
\end{lstlisting}
  \caption{Successful account response retrieved (Action \textit{A4.2})}\label{fig:AccessTokenA4.2]}
\end{figure}

The data provided by the endpoint illustrates account information retrieved from the API, showing the structure of financial accounts. Each account within the Account array includes \texttt{AccountId}, currency, account type, account subtype, description, and nickname, along with additional attributes such as scheme name, identification, and user name.

\subsection{Request Transaction}
The process of requesting a transaction is similar to requesting account information, with the same parameters but involving different endpoints: \smallurl{https://ob.sandbox.natwest.com/open-banking/v3.1/aisp/accounts/{{accountIdUsedToRequestTransactions}}/transactions}. Both actions typically involve initiating a request via an API, obtaining authorisation (often through consent), and using specific endpoints to execute the desired action.

\begin{figure}[H]
  \centering
  \begin{lstlisting}[language=json]
{
    "Data": {
        "Transaction": [
            {
                "AccountId": "918da596-c50f-499f-bea7-2b77adcb96f3",
                "TransactionId": "7626d3ef-eb01-4336-ad44-a9710c085328",
                "CreditDebitIndicator": "Credit",
                "Status": "Booked",
                "BookingDateTime": "2024-03-29T11:58:00.000Z",
                "Amount": {
                    "Amount": "250.00",
                    "Currency": "GBP"
                },
                "ProprietaryBankTransactionCode": {
                    "Code": "TFR"
                },
                "TransactionInformation": "Monthly savings",
                "Balance": {
                    "CreditDebitIndicator": "Credit",
                    "Type": "Information",
                    "Amount": {
                        "Amount": "125680.92",
                        "Currency": "GBP"
                    }
                }
            }
        ]
    },
    "Links": {
        "Self": "https://ob.sandbox.natwest.com/open-banking/v3.1/aisp/accounts/918da596-c50f-499f-bea7-2b77adcb96f3/transactions?page=0",
        "First": "https://ob.sandbox.natwest.com/open-banking/v3.1/aisp/accounts/918da596-c50f-499f-bea7-2b77adcb96f3/transactions?page=0",
        "Last": "https://ob.sandbox.natwest.com/open-banking/v3.1/aisp/accounts/918da596-c50f-499f-bea7-2b77adcb96f3/transactions?page=0"
    },
    "Meta": {
        "TotalPages": 1
    }
}
\end{lstlisting}
  \caption{Successful transaction response retrieved (Action \textit{A4.2})}\label{fig:AccessTokenA4.2t}
\end{figure}

This process is essential for enabling the AISP to provide financial services based on PSU account data, while adhering to Open Banking regulations and standards.

The post-authorisation endpoint provides a comprehensive list of all accounts for exploration. Each account is detailed with essential attributes like \texttt{AccountId}, \texttt{Currency}, and \texttt{AccountType}, allowing for detailed analysis of financial holdings. According to the response shown in Figure \ref{fig:AccessTokenA4.2t}, the PSU has two distinct accounts.

\end{document}